\documentclass[english,aps,prl,amsmath,twocolumn,superscriptaddress]{revtex4-1}

\usepackage{graphicx}
\usepackage{bm}
\usepackage[usenames,dvipsnames,svgnames,table]{xcolor}
\usepackage[colorlinks=true,linkcolor=RoyalBlue,citecolor=RoyalBlue]{hyperref}
\usepackage{babel}

\begin{document}

\title{Nonreciprocal phonon dichroism induced by Fermi pocket anisotropy in 2D Dirac materials}

\author{Wen-Yu Shan}
\email{wyshan@gzhu.edu.cn}
\affiliation{Department of Physics, School of Physics and Materials Science, Guangzhou University, Guangzhou 510006, China}

\date{\today}

\begin{abstract}

Electrons in two-dimensional (2D) Dirac materials carry local band geometric quantities, such as Berry curvature and orbital magnetic moment, which, combined with electron-phonon coupling, may affect phonon dynamics in unusual way. Here, we propose intrinsic nonreciprocal linear and circular phonon dichroism in magnetic 2D Dirac materials, which originate from nonlocal band geometric quantities of electrons and reduce to pure Fermi-surface properties for acoustic phonons. We find that to acquire the nonreciprocity, the Fermi pocket anisotropy rather than the chirality of electrons is crucial. Two possible mechanisms of Fermi pocket anisotropy are suggested: (i) trigonal warping and out-of-plane magnetization or (ii) Rashba spin-orbit interaction and in-plane magnetization. As a concrete example, we predict appreciable and tunable nonreciprocal phonon dichroism in 2H-MoTe$_2$ on EuO substrate. Our finding points to a different route towards electrical control of phonon nonreciprocity for acoustoelectronics applications based on 2D quantum materials.

\end{abstract}

\maketitle

$\textit{Introduction}$.---Electrons in two-dimensional (2D) Dirac materials, including graphene~\cite{neto2009} and transition metal dichalcogenides~\cite{xiao2012}, carry local band geometric quantities, such as Berry curvature~\cite{berry1984,haldane1988,xiao2010} and orbital magnetic moment~\cite{xiao2007,yao2008}, as a result of broken time-reversal or inversion symmetry. These geometric quantities have recently been recognized to play pivotal roles in various electronic phenomena. When further taking into account the electron-phonon (e-ph) coupling, the geometric information of electrons can be inherited by phonons, leading to unusual phononic behaviors, e.g., phonon Hall viscosity~\cite{barkeshli2012}, anomalous phonon effective charge~\cite{rinkel2017,ren2021}, phonon magnetochiral effect~\cite{spivak2016,sengupta2020} and phonon helicity~\cite{hu2021}.

On the other hand, phonon nonreciprocity, referring to the asymmetric phononic behaviors upon reversing the propagation vector $\bm q$, are attractive for phononics applications~\cite{li2012}. The physical origins of phonon or surface-acoustic-wave nonreciprocity have been ascribed to Lorentz force~\cite{heil1982,liu2020}, asymmetric magnon-phonon coupling or magnon dispersion~\cite{emtage1976,sasaki2017,nomura2019,xu2020,shanp2020,hernandez2020,tateno2020}, nonlinearity~\cite{liang2010} and parametric time dependence~\cite{xu2019}. In contrast, electrically controlled phonon nonreciprocity based on e-ph coupling has been less explored~\cite{sengupta2020}. One possible reason is that such phenomena of nonreciprocity are generally weak in bulk materials. Given the enhanced e-ph interaction and highly-tunable electronic band structure~\cite{ge2013,navarro2016,choi2018,sohier2019,han2021}, 2D materials seem to be promising in the pursuit of e-ph coupling-driven nonreciprocity.

\begin{figure}[b]
\centering \includegraphics[width=0.49\textwidth]{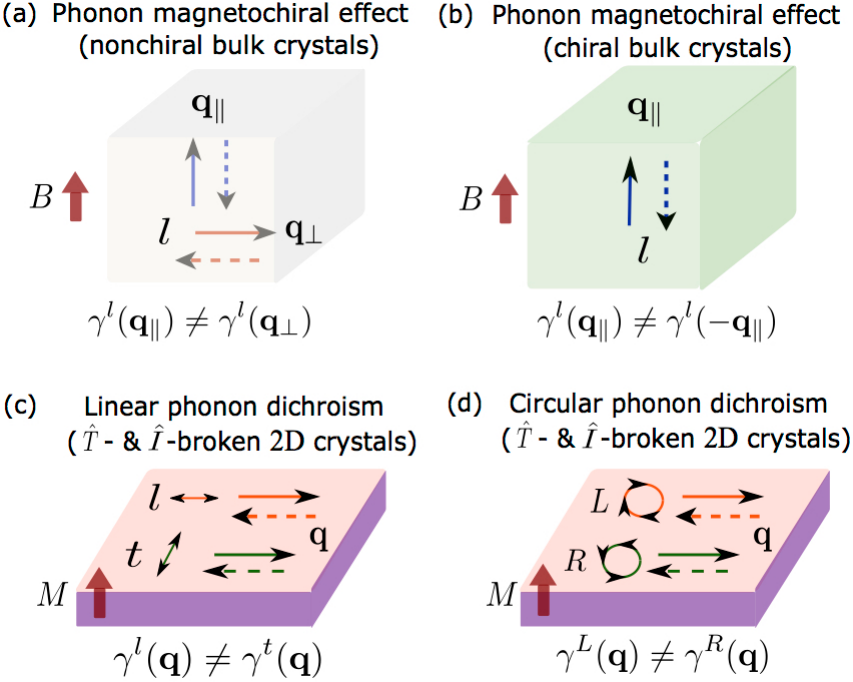}
\caption{ Schematics of phonon magnetochiral effect for (a) nonchiral and (b) chiral bulk crystals, (c) linear and (d) circular phonon dichroism for $\hat{T}$- and $\hat{I}$-broken 2D crystals. $\gamma^i(\bm q)$ are the phonon absorption coefficients, with phonon wave vector $\bm q$ and polarization $i$. In (a)-(b), $\bm q_{\parallel}$ ($\bm q_{\bot}$) is parallel (perpendicular) to the external magnetic field $B$. In (c)-(d), $\bm q$ is in plane while magnetization $M$ is out of plane. $l/t$ labels phonons with longitudinal (transverse) polarization, and $L/R$ labels left- (right-) handed circular phonons. The polarization is indicated by double arrows in (c)-(d). Dash lines refer to the reversal process when changing $\bm q$ to $-\bm q$. }
\label{fig:phonon_nonrec}
\end{figure}

\begin{table*}[tp]
\caption{ Chirality and reciprocity for phonon magnetochiral effect (PMC), linear phonon dichroism (LPD) and circular phonon dichroism (CPD). $H_{0}$, $H_Z$, $H_X$, $H_R$ and $H_w$ are the Hamiltonian for intrinsic monolayer transition metal dichalcogenides, out-of-plane magnetization, in-plane (zigzag-direction) magnetization, Rashba SOI and trigonal warping, respectively.}
\label{tab:nonreciprocity}%
\begin{ruledtabular}
\begin{tabular}{*4c} 
 & chiral & reciprocal  & reciprocal \\  
 &  & $(\bm q\rightarrow-\bm q)$  & $(\bm B/\bm M\rightarrow-\bm B/\bm M)$\\ \hline
PMC (bulk)~\cite{spivak2016} & no & yes & yes \\ 
PMC (bulk)~\cite{sengupta2020} & yes & no & no \\ 
LPD/CPD ($H_{0}+H_{Z}$)~\cite{shan2022}   & no & yes & yes (LPD), no (CPD) \\
LPD/CPD ($H_{0}+H_{Z}+H_{R}$)~\cite{shan2022} & yes & yes & yes (LPD), no (CPD)\\
LPD/CPD ($H_{0}+H_Z+H_w$) & no & no & no (LPD), no (CPD) \\
LPD/CPD ($H_{0}+H_X$) & no & yes & yes (LPD), no (CPD) \\
LPD/CPD ($H_{0}+H_X+H_R$) & no & no & no (LPD), no (CPD) \\
\end{tabular}
\end{ruledtabular}
\end{table*}

An attempt has been made to study the circular phonon dichroism (CPD) in monolayer transition metal dichalcogenides~\cite{shan2022}, where there is a difference of absorption coefficients $\gamma$ between left ($L$)- and right ($R$)-handed circularly polarized phonons [see Fig.~\ref{fig:phonon_nonrec} (d)]. Based on symmetry principles~\cite{szaller2013,cheong2018,tokura2018}, such CPD was expected to be nonreciprocal, i.e., distinct $\gamma$ when reversing $\bm q$ to $-\bm q$, as a result of the simultaneously broken time-reversal $\hat{T}$ and space-inversion symmetry $\hat{I}$. However, analytical calculations suggested unexpected reciprocal behaviors $\gamma^{L/R}(\bm q)=\gamma^{L/R}(-\bm q)$, and nonreciprocal behaviors $\gamma^{L/R}(\bm M)\neq\gamma^{L/R}(-\bm M)$ under the reversal of magnetization $\bm M$. More surprisingly, when further incorporating the Rashba spin-orbit interaction (SOI), the reciprocity in phonon attenuation between $\pm\bm q$ survives despite the fact that the system becomes chiral due to the lack of mirror planes. This is in stark contrast to the phonon magnetochiral effect in bulk crystals [see Fig.~\ref{fig:phonon_nonrec} (a)-(b)], which shows nonreciprocal (reciprocal) behaviors when the underlying system is chiral (nonchiral)~\cite{spivak2016,sengupta2020}. Here we adopt the concept of chirality from chiral crystals~\cite{nomura2019,yu2023}, that is, a system is $chiral$ when distinguishable from its mirror image. From the viewpoint of symmetries, an electron acquires chirality when all mirror symmetries and time-reversal symmetry are broken, otherwise it is nonchiral. The different nonreciprocal behaviors between CPD and phonon magnetochiral effect raise a question: What kind of geometric information (beyond the chirality) of electrons is inherited by phonons and induce the phonon nonreciprocity? To address it, a deeper understanding of the interplay of electronic band geometry, symmetry and phonon nonreciprocity is required.

In this paper, we show that the linear and circular phonon dichroism arise from nonlocal band geometric quantities of electrons in magnetic 2D Dirac materials. For acoustic phonons, these effects reduce to pure Fermi-surface properties. To acquire the nonreciprocity, we find that the chirality is not a necessary condition [as summarized in Table \ref{tab:nonreciprocity}], whereas the anisotropy of each Fermi pocket is crucial, particularly in multivalley systems. We propose two possible mechanisms to realize the Fermi pocket anisotropy [as shown in Table \ref{tab:nonreciprocity}]: (i) trigonal warping and out-of-plane magnetization or (ii) Rashba SOI and in-plane magnetization. To demonstrate our theory, nonreciprocal LPD and CPD are discussed in 2H-MoTe$_2$ deposited on EuO substrate. Our study uncovers a connection between electronic band geometry and nonreciprocal phonon attenuation, and paves the way towards electrical control of phonon nonreciprocity for acoustoelectronics applications.

$\textit{Linear and circular phonon dichroism}$.---The Hamiltonian reads $\hat{\mathcal{H}}=\hat{\mathcal{H}}_e+\hat{\mathcal{H}}_{e-ph}$, where $\hat{\mathcal{H}}_e$ and $\hat{\mathcal{H}}_{e-ph}$ correspond to electronic and e-ph coupling term, respectively. The general form of e-ph coupling follows $\hat{\mathcal{H}}_{e-ph}=\sum_{\bm k,\bm q}\psi^+(\bm k)[\bm u(\bm q)\cdot\hat{\bm T}(\bm q)]\psi(\bm k-\bm q)$~\cite{landau1959,suzuura2002,liu2017,shan2020,shan2022}, where $\psi(\bm k)$ and $\psi^+(\bm k)$ are the annihilation and creation operator of electrons. For acoustic phonons, $\bm u(\bm q)$ is a Fourier transform of in-plane collective displacement $\bm u(\bm r)=(u_x,u_y)$ from the equilibrium position of ions, with phonon propagation vector $\bm q$. $\hat{\bm T}(\bm q)$ can be regarded as a force operator acting on ions exerted by electrons. 

For multivalley systems, the valley-resolved retarded response function reads~\cite{supple}
\begin{equation}
\begin{split}
&\chi_{\alpha\beta}^{\tau}(\bm q,\omega)=
\sum_{n,m}\int\frac{\hbar d^2\bm k}{\rho(2\pi)^2}
F_{nm}^{\tau}(\omega,\bm k,\bm q)[S_{\alpha\beta}^{\tau}(\omega,\bm k,\bm q)]_{mn},
\end{split}
\end{equation}
where $\alpha,\beta=x,y$, $\tau=\pm1$ labels valley $K_{\pm}$, $\rho$ is the 2D mass density, and $m,n$ are band indices. The dynamical factor 
\begin{equation}
\begin{split}
F_{nm}^{\tau}(\omega,\bm k,\bm q)&=\frac{f_{\tau,m,\bm k}-f_{\tau,n,\bm k'}}{\omega+i\delta+E_{\tau,m,\bm k}-E_{\tau,n,\bm k'}}
\end{split}
\end{equation}
and geometric factor
\begin{equation}
\begin{split}\label{geometric}
[S_{\alpha\beta}^{\tau}(\omega,\bm k,\bm q)]_{mn}&=
\langle\psi_{\tau,m,\bm k}|\hat{T}_{\alpha}^{\tau}(\bm q)|\psi_{\tau,n,\bm k'}\rangle\\
&\times\langle\psi_{\tau,n,\bm k'}|\hat{T}_{\beta}^{\tau}(-\bm q)|\psi_{\tau,m,\bm k}\rangle,
\end{split}
\end{equation}
where $E_{\tau,m,\bm k}$ and $|\psi_{\tau,m,\bm k}\rangle$ are eigen-dispersion and wave functions of $\hat{\mathcal{H}}_e$, respectively. $\bm k'=\bm k-\bm q$. $f_{\tau,m,\bm k}$ ($f_{\tau,n,\bm k'}$) is the Fermi distribution function and $\delta$ is a positive infinitesimal. $\hat{\bm T}^{\tau}(\bm q)$ is the valley-dependent force operator.

We mainly focus on the anti-Hermitian part of $[\chi_{\alpha\beta}]$ matrix, that is, $-2i\omega\gamma(\bm q,\omega)$, which physically describes the dissipation of phonons. The form of the Hermitian matrix $\gamma(\bm q,\omega)$ reads~\cite{supple}
\begin{equation}
\begin{split}
\gamma(\bm q,\omega) = 
\left[\begin{array}{cc}
\gamma_{D}+\gamma_{\bar{D}} & \gamma_{\bar{A}}+i\gamma_{A} \\
\gamma_{\bar{A}}-i\gamma_{A} & \gamma_{D}-\gamma_{\bar{D}} \\
\end{array}\right],
\end{split}
\end{equation}
$\gamma_{D}=(\gamma_{xx}+\gamma_{yy})/2$ and $\gamma_{\bar{D}}=(\gamma_{xx}-\gamma_{yy})/2$ refer to the symmetric and antisymmetric normal (longitudinal) absorption. $\gamma_{\bar{A}}=\mathrm{Re}[\gamma_{xy}]$ and $\gamma_{A}=\mathrm{Im}[\gamma_{xy}]$ refer to the symmetric and antisymmetric anomalous (Hall) absorption. When linearly polarized longitudinal or transverse phonons are injected, i.e., $|u_{l}(\bm q)\rangle=[\cos\phi_{\bm q},\sin\phi_{\bm q}]^T$ and $|u_{t}(\bm q)\rangle=[-\sin\phi_{\bm q},\cos\phi_{\bm q}]^T$, the damping (absorption) coefficients $\gamma^{l/t}=\gamma_D\pm\cos2\phi_{\bm q}\gamma_{\bar{D}}\pm\sin2\phi_{\bm q}\gamma_{\bar{A}}$, where the azimuthal angle $\phi_{\bm q}=\tan^{-1}(q_y/q_x)$. The difference between $\gamma^l$ and $\gamma^t$ defines the $linear$ $phonon$ $dichroism$ (LPD) [see Fig.~\ref{fig:phonon_nonrec} (c)]. On the other hand, when circularly polarized phonons are considered, $|u_{L/R}(\bm q)\rangle=\frac{1}{\sqrt{2}}[1,\pm i]^T$, the damping coefficients $\gamma^{L/R}=\gamma_{D}\mp\gamma_{A}$. The difference between $\gamma^L$ and $\gamma^R$ defines the $circular$ $phonon$ $dichroism$ (CPD) [see Fig.~\ref{fig:phonon_nonrec} (d)]. As LPD and CPD effects arise from different parts of $\gamma$ matrix, the two phenomena may behave rather differently [see Table \ref{tab:nonreciprocity}].

\begin{figure}[t]
\centering \includegraphics[width=0.49\textwidth]{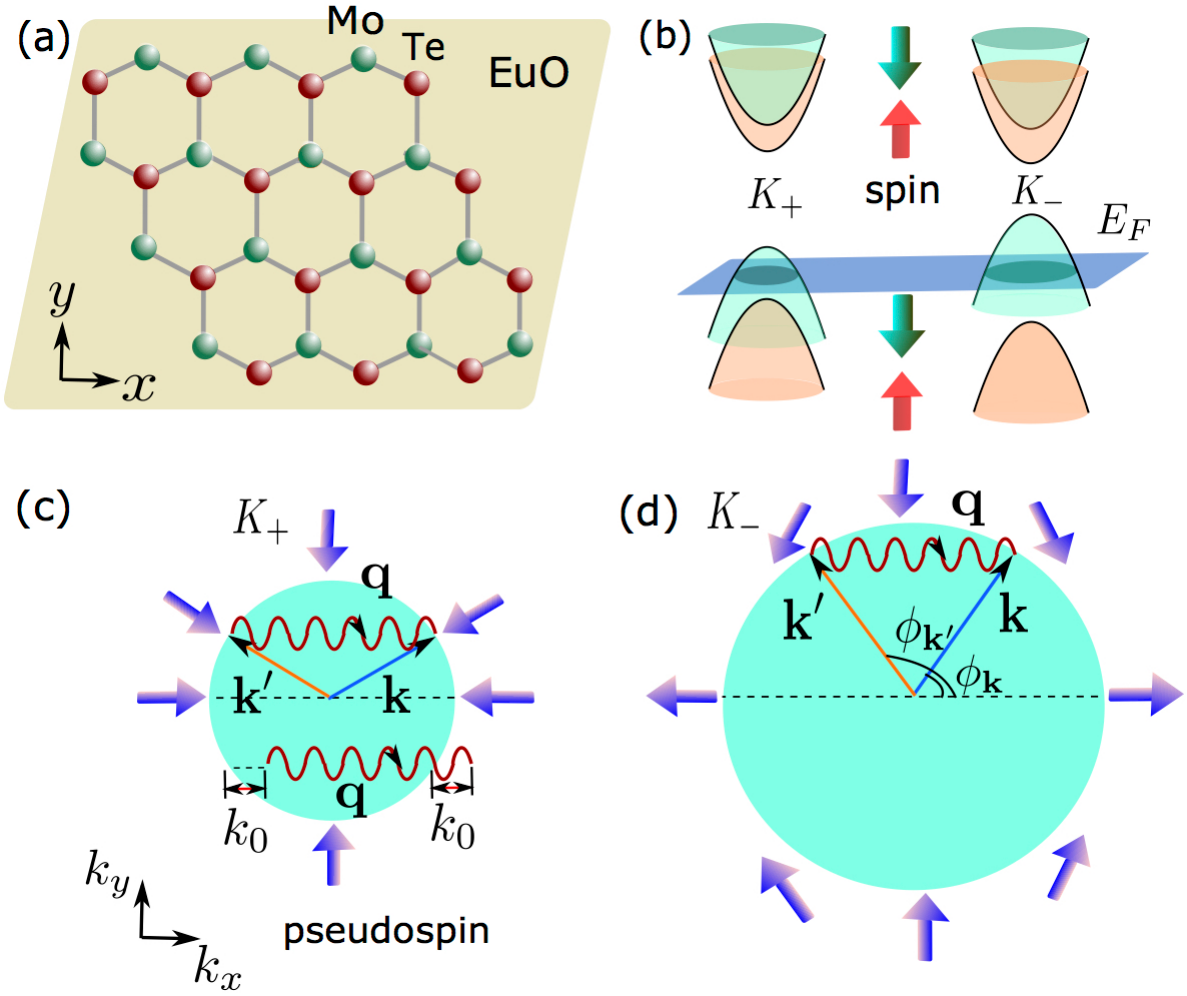}
\caption{ Schematics of (a) the lattice and (b) electronic band structure of 2H-MoTe$_2$ on EuO substrate. Pseudospin structure on the Fermi surface (intersected by $E_F$ in (b)) at (c) valley $K_+$ and (d) valley $K_-$. In (b), red (green) arrows label spin-up (-down) bands. In (c) and (d), purple arrows denote the in-plane projection of pseudospin for states lying on the Fermi surface. Phonon wave vector $\bm q$ is assumed to be along the $\hat{x}$ (zigzag) direction, connecting two electronic states with wave vector $\bm k$ and $\bm k'=\bm k-\bm q$. $k_0$ is a small deviation of wave vector from the Fermi surface due to energy offset $\omega$.}
\label{fig:phonon_geometric}
\end{figure}

$\textit{Geometric origin}$.---We now reveal the geometric origin of LPD and CPD effects. For 2D materials, acoustic phonon energy $\omega$ is usually much smaller than the electronic Fermi energy $E_F$. In this sense, in the low-temperature limit, the dynamical factor reduces to 
\begin{equation}
\begin{split}\label{constraint}
\mathrm{Im}[F^{\tau}_{nm}]\approx-\pi\omega\delta(E_{\tau,m,\bm k}-E_F)\delta(\omega+E_{\tau,m,\bm k}-E_{\tau,n,\bm k'}).
\end{split}
\end{equation}
Physically, this means that acoustic phonons only introduce transitions between electrons on the Fermi surface. If a single band is intersected by the Fermi level at each valley [see Fig.~\ref{fig:phonon_geometric} (b)], only intraband transition $m=n$ is allowed.

The geometric factor $(S_{\alpha\beta}^{\tau})_{mn}$ characterizes a connection between states with wave vector $\bm k$ and $\bm k'$, triggered by the force operator $\hat{\bm T}^{\tau}(\bm q)$. Here $\bm k$ and $\bm k'=\bm k-\bm q$ are not necessarily close [see Fig.~\ref{fig:phonon_geometric} (c) and (d)], suggesting that $(S_{\alpha\beta}^{\tau})_{mn}$ has a $nonlocal$ nature in the momentum space. This is in stark contrast to the optical~\cite{xiao2010} or phononic response in the long-wavelength limit~\cite{sengupta2020,hu2021}, where the occurrence of Berry curvature and orbital magnetic moment all relies on the local operations in the momentum space. As a comparison, the long-wavelength limit of $(S_{\alpha\beta}^{\tau})_{mn}$ is also discussed in the Supplementary Material~\cite{supple}.

The role of $\hat{\bm T}^{\tau}(\bm q)$ in Eq. (\ref{geometric}) can be viewed as a rotation of (pseudo-) spins of Bloch states. For conventional deformation potentials $\hat{\bm T}^{\tau}(\bm q)=ig\bm q$, $\chi_{\alpha\beta}$ reduces to normal electron polarization function~\cite{mahan2000,giuliani2005}, corresponding to a vanishing CPD signal. In 2D hexagonal Dirac crystals, the e-ph coupling behaves more like pseudo-gauge potentials~\cite{suzuura2002,bhalla2022}, e.g., $\hat{\bm T}^{\tau=-1}(\bm q)=ig[\bm q\cdot\bm\sigma,(\bm q\times\bm\sigma)_z]$ and $\hat{\bm T}^{\tau=1}(\bm q)=\mathcal{K}[\hat{\bm T}^{\tau=-1}(-\bm q)]$, with the complex conjugation $\mathcal{K}$. Here $\hat{T}_x^{\tau}(\bm q)$ and $\hat{T}_y^{\tau}(\bm q)$ act as a rotation of angle $\pi$ in the (pseudo-) spin subspace about the axis along the direction $(q_x,-\tau q_y)$ and $(\tau q_y,q_x)$, respectively. In this sense, $\chi_{\alpha\beta}$ can be regarded as a peculiar type of dynamical spin susceptibility~\cite{mahan2000,giuliani2005}, which, however, only depends on states on the Fermi surface. 

$\textit{Model}$.---To facilitate the discussion, we consider a generic model for magnetic 2D Dirac materials: $\hat{\mathcal{H}}_{e}=\sum_{\bm k}\psi^+(\bm k)H_e(\bm k)\psi(\bm k)$, with $H_e(\bm k) = H_0+H_{\bm n}$ and
\begin{equation}
\begin{split}\label{basic}
&H_0 = \frac{\Delta}{2}\sigma_z + \hbar v(\tau\sigma_xk_x+\sigma_yk_y)
+ \tau s_z(\lambda_c\sigma_++\lambda_v\sigma_-),\\
&H_{\bm n}  = -\bm s\cdot\bm n(M_c\sigma_++M_v\sigma_-).
\end{split}
\end{equation}
$\bm s$ and $\bm \sigma$ are Pauli matrices acting on spin and pseudospin (orbital) subspace, respectively. $\sigma_{\pm}=\frac{1}{2}(\sigma_0\pm\sigma_z)$. $H_{\bm n}$ refers to the magnetic exchange coupling. $\lambda_{c/v}$ and $M_{c/v}$ characterize the intrinsic SOI and Zeeman field in the sublattice space, respectively. This model can be applied to graphene~\cite{kane2005}, monolayer transition metal dichalcogenides~\cite{qi2015,norden2019} and other 2D Dirac materials~\cite{cui2021}. Here we consider an out-of-plane magnetization $H_Z\equiv H_{\bm n=\hat{z}}$ and set the Fermi level in the valence band, whose dispersion is shown schematically in Fig. \ref{fig:phonon_geometric} (b). 

We find that $\gamma_A^{\tau}=0$ exactly on the Fermi surface, whereas $\gamma_{D/\bar{D}/\bar{A}}^{\tau}$ are generally nonzero~\cite{supple}. The vanishing $\gamma_A^{\tau}$ is a result of Dirac model rather than specific symmetries. This is in contrast to the non-vanishing optical conductivities of Dirac model. Such difference originates from the different optical and phononic processes. For optical processes, Hall conductivities are contributed by the band geometric quantities of both conduction and valence bands since the interband transitions of electrons are optically induced. For acoustic phonons, electronic transitions are within valence bands with different wave vectors since phonons are unable to induce the transitions between conduction and valence bands [see Fig. \ref{fig:phonon_geometric} (b)]. As a result, $\gamma_A^{\tau}$ depends on the information of band geometries of purely valence bands, and vanishes for Dirac model. The vanishing $\gamma_A^{\tau}$, however, does not imply that CPD vanishes since there is a small energy offset $\omega$ in $\mathrm{Im}[F^{\tau}_{vv}]$ which drives electrons slightly away from the Fermi surface (that is, $k_0$ in Fig. \ref{fig:phonon_geometric} (c)). As a result, we find that $\mathrm{Im}[S_{xy}^{\tau}]_{vv}\propto\omega$~\cite{supple}, which gives rise to a finite, linear-in-$\omega$ contribution to $\gamma_A^{\tau}$.

$\textit{Nonreciprocity and symmetry}$.---When reversing the phonon propagation vector $\bm q$, we find that $\gamma^i(\bm q,\omega)=\gamma^i(-\bm q,\omega)$, with $i=l,t,L,R$, indicating that both LPD and CPD are reciprocal [see Fig. \ref{fig:phonon_nonrec} (c) and (d)]. To understand it, we check the remaining symmetries for electronic Hamiltonian $H_e(\bm k)=H_0+H_Z$ at each valley~\cite{supple}: mirror reflection $\hat{\sigma}_h$ $(z\rightarrow-z)$, three-fold rotation $\hat{C}_3$, and two hidden ``inversion" symmetries $\hat{\mathcal{P}}_1=s_0\sigma_z$ and $\hat{\mathcal{P}}_2=s_z\sigma_z$. Note that only $\hat{\mathcal{P}}_1$ and $\hat{\mathcal{P}}_2$ symmetries relate electrons with $\pm\bm k$ at a single valley $K_{\tau}$, that is, $\hat{\mathcal{P}}_{1(2)}^+H_e(\bm k)\hat{\mathcal{P}}_{1(2)}=H_e(-\bm k)$. By contrast, $\hat{\sigma}_h$ relates $(\tau,\bm k)$ to itself and $\hat{C}_3$ relates $(\tau,\bm k)$ to $(\tau,\hat{C}_3\bm k)$, neither of which is relevant to the reciprocal relations between $\pm\bm q$. Here $\bm k$ is a relative wave vector away from the valley center. Furthermore, we find that $\hat{\mathcal{P}}_{1(2)}$ transforms the force operator $\hat{\bm T}^{\tau}(\bm q)$ by $\hat{\mathcal{P}}_{1(2)}^+\hat{\bm T}^{\tau}(\bm q)\hat{\mathcal{P}}_{1(2)}=\hat{\bm T}^{\tau}(-\bm q)$. As a result, we recognize that either $\hat{P}_1$ or $\hat{P}_2$ symmetry is the origin of reciprocal relations for phonon attenuation between $\pm\bm q$~\cite{supple}.

\begin{figure}[t]
\centering \includegraphics[width=0.50\textwidth]{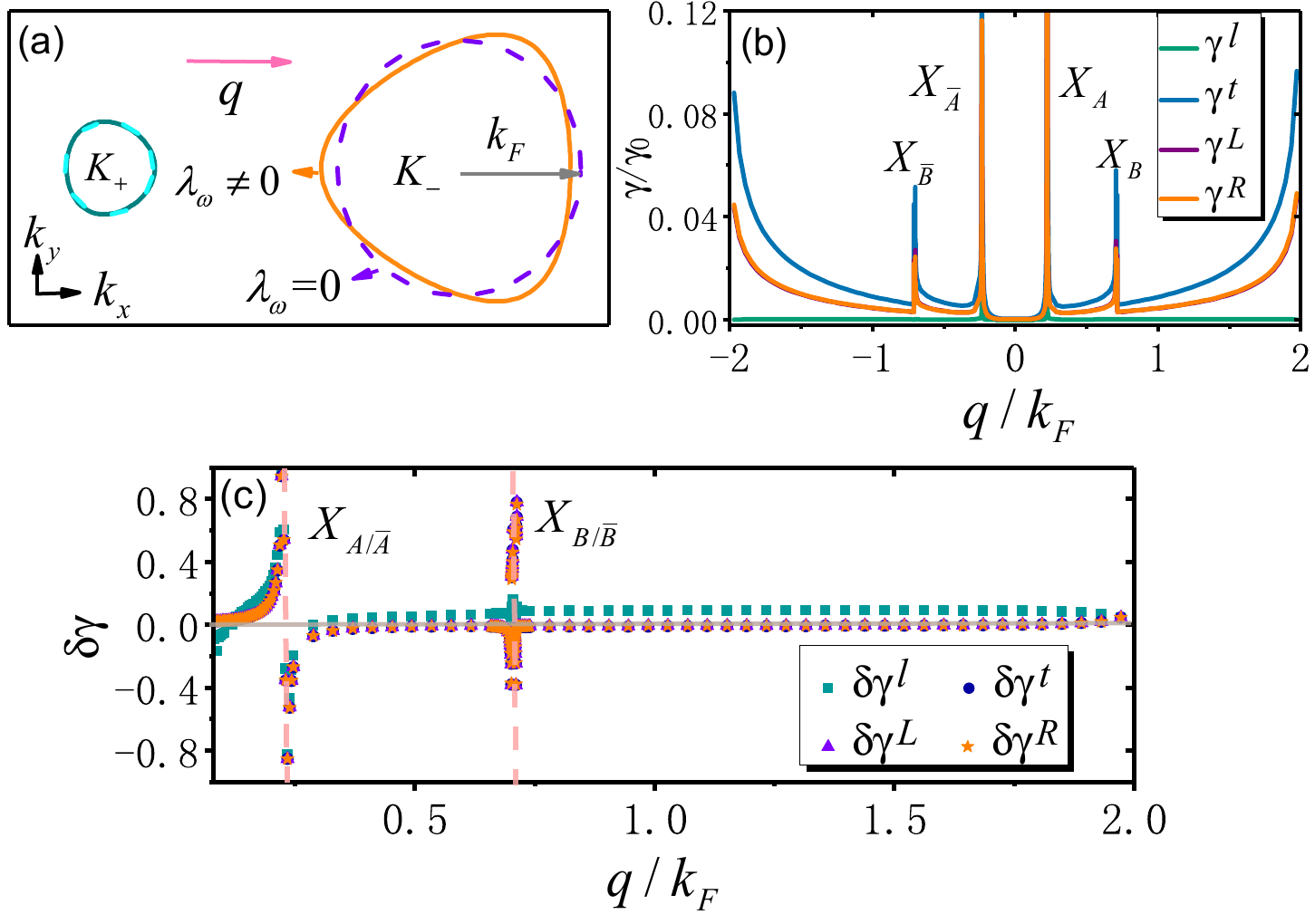}
\caption{ (a) Anisotropic (Isotropic) Fermi pockets of monolayer MoTe$_2$ with (without) trigonal warping, denoted by solid (dashed) lines and $\lambda_{w}\neq0$ ($\lambda_{w}=0$). $k_F$ is the Fermi wave vector at valley $K_-$ without trigonal warping. $\bm q=q\hat{x}$ and $\bm M=M\hat{z}$. (b) Linear and circular phonon dichroism $\gamma^{l/t}$ and $\gamma^{L/R}$ (in units of $\gamma_0$) as functions of phonon wave vector $q$ (in units of $k_F$). (c) Degree of nonreciprocity $\delta\gamma^i$ (with $i=l,t,L,R$) as functions of phonon wave vector $q$. Parameters: $\Delta=1.05$ eV, $\lambda_v=0.11$ eV, $\lambda_c=0.029$ eV, $\hbar v=2.33$ eV$\cdot\AA$, $M_c=0.206$ eV, $M_v=0.17$ eV~\cite{qi2015}, $\lambda_{w}=-1.0$ eV$\cdot\AA^2$, $E_F=-0.5$ eV, $g=0.32$ eV~\cite{chen2016}, $\gamma_0=2.23\times10^9$ s$^{-1}$~\cite{shan2022}. Phonon dispersion $\omega=\hbar c_{l/t}|\bm q|$, with longitudinal (transverse) sound velocity $c_l=3.64\times10^3$ m/s ($c_t=2.21\times10^3$ m/s)~\cite{rano2020}.}
\label{fig:nonrec_warp}
\end{figure}

The role of chirality can be examined by incorporating Rashba SOI 
\begin{equation}
\begin{split}
H_R=\lambda_R(\tau s_y\sigma_x-s_x\sigma_y)
\end{split}
\end{equation}
into $H_e(\bm k)$. Such term naturally breaks $\hat{\sigma}_h$ symmetry, meaning that no mirror plane exists. Given that time-reversal symmetry is already broken in Eq. (\ref{basic}), the system becomes $chiral$. Detailed calculations show that LPD and CPD are still reciprocal [see Table \ref{tab:nonreciprocity}], thereby excluding a possible origin of nonreciprocity due to nonzero chirality. On the other hand, such result can be explained by recognizing that an introduction of $H_R$ still preserves $\hat{\mathcal{P}}_2$ symmetry, which establishes a link between $\pm\bm k$.

The role of $\hat{C}_3$ symmetry can be further examined by replacing the out-of-plane magnetization $H_Z$ with the in-plane magnetization $H_X\equiv H_{\bm n=\hat{x}}$. We find that LPD and CPD are reciprocal [see Table \ref{tab:nonreciprocity}] despite the fact that $H_X$ breaks the $\hat{C}_3$ symmetry. This implies that $\hat{C}_3$ symmetry is also not the origin of nonreciprocal behaviors. By contrast, $H_X$ preserves $\hat{\mathcal{P}}_1$ symmetry~\cite{supple}, which still provides a link between $\pm\bm k$.

Therefore to acquire the nonreciprocity, the simultaneous $\hat{\mathcal{P}}_1$ and $\hat{\mathcal{P}}_2$ symmetry breaking is required. This naturally leads to an anisotropy of Fermi pocket at each valley. To realize it, two possible schemes are proposed in the following.  Moreover, the nonreciprocal responses under the reversal of magnetization $\bm M\rightarrow-\bm M$ can be derived by the generalized Onsager reciprocal theorem $\gamma(\bm q,\bm M)=\gamma^T(-\bm q,-\bm M)$~\cite{onsager1931,rikken2001,tokura2018,supple} [see Table \ref{tab:nonreciprocity}].

\begin{figure}[t]
\centering \includegraphics[width=0.50\textwidth]{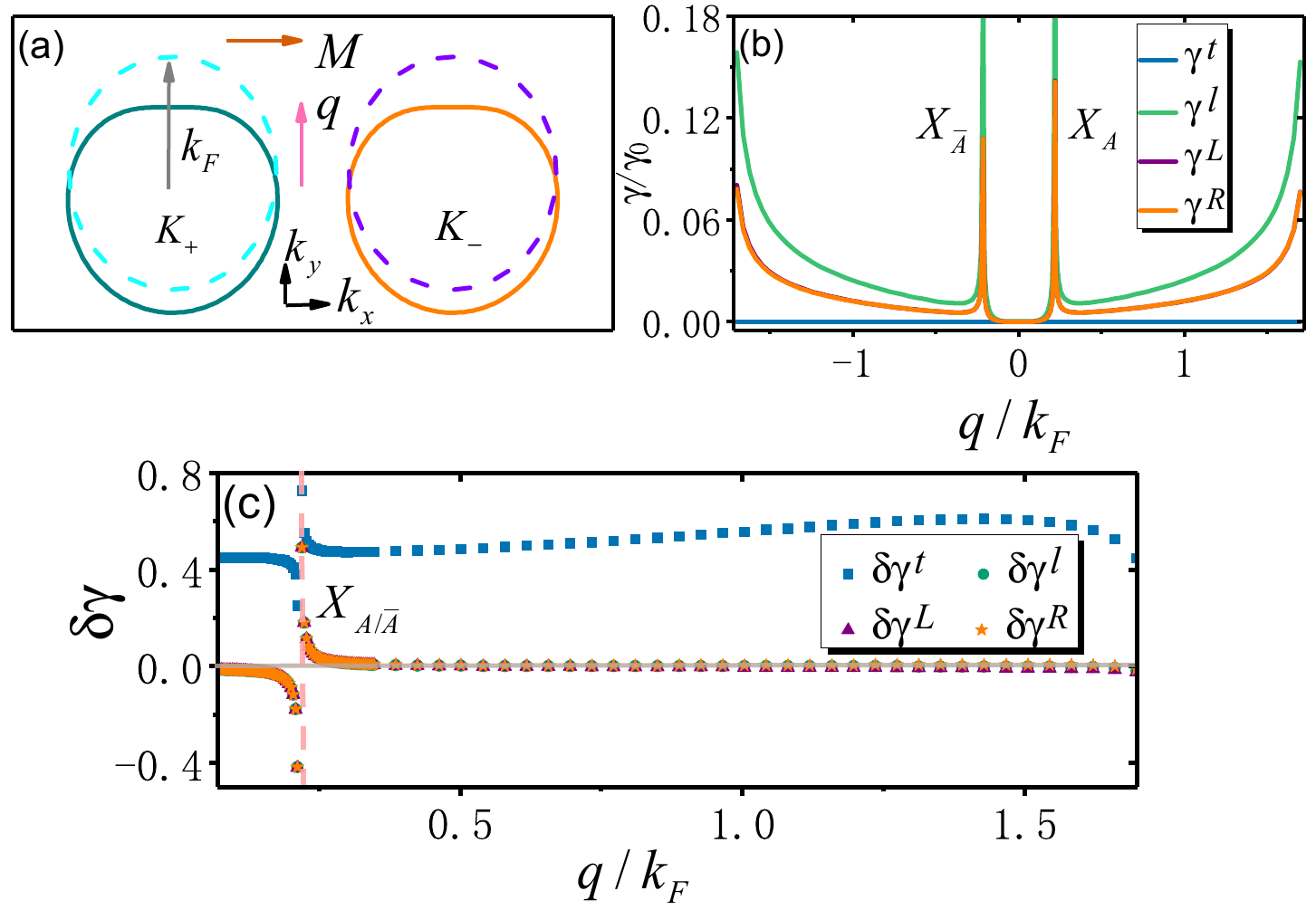}
\caption{ (a) Anisotropic (Isotropic) Fermi pockets of monolayer MoTe$_2$ with Rashba SOI in the presence (absence) of in-plane magnetization, denoted by solid (dashed) lines. $k_F$ is the Fermi wave vector at valley $K_{\pm}$ without magnetization. $\bm q=q\hat{y}$ and $\bm M=M\hat{x}$. (b) Linear and circular phonon dichroism $\gamma^{l/t}$ and $\gamma^{L/R}$ (in units of $\gamma_0$) as functions of phonon wave vector $q$ (in units of $k_F$). (c) Degree of nonreciprocity $\delta\gamma^i$ as functions of phonon wave vector $q$. Parameters: $\lambda_{R}=0.2$ eV, $E_F=-0.6$ eV.}
\label{fig:nonrec_rashba}
\end{figure}

$\textit{Trigonal warping and out-of-plane magnetization}$.---One scheme to acquire the nonreciprocity is to consider the Hamiltonian $H_e(\bm k)=H_0+H_w+H_Z$, with trigonal warping term~\cite{liu2013,kormanyos2013,battilomo2019}
\begin{equation}
\begin{split}
H_{w}=\lambda_{w}[(k_x^2-k_y^2)\sigma_x-2\tau k_xk_y\sigma_y].
\end{split}
\end{equation}
This leads to the Fermi pocket anisotropy at each valley, as shown in Fig. \ref{fig:nonrec_warp} (a). The $q$-dependent $\gamma^{l/t}$ and $\gamma^{L/R}$ exhibit two peaks for both positive and negative $q$ [see Fig. \ref{fig:nonrec_warp} (b)], labeled by $X_{A/\bar{A}}$ and $X_{B/\bar{B}}$. These peaks have distinct physical origins. Peaks $X_{A/\bar{A}}$ at $q/k_F=\pm0.22$ arise from the divergent $[dE_{\tau=-1,v,k,k'}/dk]^{-1}$ (valley $K_-$ is dominant over $K_+$), which can be regarded as a new type of joint density of states between wave vector $\bm k$ and $\bm k-\bm q$ of valence band. This is a natural consequence of Fermi pocket anisotropy. On the other hand, peaks $X_{B/\bar{B}}$ at $q/k_F=\pm0.70$ are due to a sudden vanishing of contributions from valley $K_+$ [see Fig. \ref{fig:nonrec_warp} (a)]. According to the dynamical factor $\mathrm{Im}[F^{\tau}_{vv}]$ from Eq. (\ref{constraint}), the critical value $q_c$ can be derived by solving the condition $\omega(q_c)+E_{\tau=1,v,\bm k}=E_{\tau=1,v,\bm k-\bm q_c}$ for valley $K_+$. When $q$ becomes larger than $q_c$, electronic scattering within valley $K_+$ by phonons is inhibited, and only scattering within valley $K_-$ exists and contributes to phonon absorption. The degree of nonreciprocity can be defined as $\delta\gamma^{i}=[\gamma^i(\bm q)-\gamma^i(-\bm q)]/[\gamma^i(\bm q)+\gamma^i(-\bm q)]$, where $i=l,t,L,R$ label different types of phonon polarization. In Fig. \ref{fig:nonrec_warp} (c), $\delta\gamma^i$ is tunable by $q$ and becomes pronounced around peaks $X_{A/\bar{A}}$ and $X_{B/\bar{B}}$. The behaviors at peaks $X_{B/\bar{B}}$ are complex due to a competition of $l$ and $t$ phonon modes, which have different critical value $q_c$.

$\textit{Rashba SOI and in-plane magnetization}$.---Another scheme to acquire the nonreciprocity is to consider the Hamiltonian $H_e(\bm k)=H_0+H_R+H_X$, whose results are shown in Fig. \ref{fig:nonrec_rashba}. Under a magnetization $\bm M$ along $x$ direction, the anisotropic Fermi pocket of each valley shifts collectively along $y$ direction [see Fig. \ref{fig:nonrec_rashba} (a)]. This excludes the possibilities of peaks $X_{B/\bar{B}}$ in $\gamma^{l/L/R}$ since the critical value $q_c$ is always the same between valley $K_+$ and $K_-$. The occurrence of peaks $X_{A/\bar{A}}$ still arises from the divergent $[dE_{\tau=\pm,v,k,k'}/dk]^{-1}$, as characteristics of Fermi pocket anisotropy. $\gamma^t$ is vanishingly small [see Fig. \ref{fig:nonrec_rashba} (b)], since $\gamma^t=\gamma_D+\gamma_{\bar{D}}\propto$ geometric factor $[S_{xx}^{\tau}]_{vv}\propto|\langle\psi_{\tau,v,\bm k}|q_y\sigma_y|\psi_{\tau,v,\bm k-\bm q}\rangle_{k_x=0}|^2$ according to Eq. (\ref{geometric}). Such matrix element almost vanishes since the pseudospin of two states $|\psi_{\tau,v,k_x=0,k_y}\rangle$ and $|\psi_{\tau,v,k_x=0,k_y-q_y}\rangle$ points oppositely along $y$ direction [see Fig. \ref{fig:phonon_geometric} (c) and (d)]. These lead to nonreciprocal LPD and CPD as shown in Fig. \ref{fig:nonrec_rashba} (c). On the other hand, when $\bm q$ is along $x$ direction, i.e., $\bm q\parallel\bm M$, the nonreciprocity vanishes.

$\textit{Discussion}$.---We have studied the nonreciprocal phonon dichroism induced by Fermi pocket anisotropy in 2D Dirac materials. We find that these effects have the origin of nonlocal electronic band geometries and are determined by the Fermi-surface properties for acoustic phonons. In multi-valley systems, the nonreciprocity is driven by the Fermi pocket anisotropy rather than the chirality. Two possible schemes are proposed to realize the Fermi pocket anisotropy in 2H-MoTe$_2$. Interestingly, similar proposals have been given in electronic systems in order to acquire the nonreciprocal current in noncentrosymmetric Rashba superconductors~\cite{wakatsuki2017,wakatsuki2018,hoshino2018}, nonreciprocal resistivity in BiTeBr~\cite{ideue2017} and unidirectional valley-contrasting photocurrent~\cite{asgari2022}. Our findings suggest that the geometric information of electrons can be inherited by phonons through the pseudo-gauge e-ph coupling, thereby inducing the phonon nonreciprocity.

In the above analysis, we have only considered the imaginary part of self-energy and neglected the real part. The treatment is appropriate since we mainly focus on the acoustic phonons, whose sound velocity is about two order of magnitude smaller than the Fermi velocity of electrons. This leads to $|\Sigma_s|/\omega\sim10^{-6}$ when $|\bm q|/k_F=0.1$; $|\Sigma_s|/\omega\sim10^{-3}$ when $|\bm q|/k_F=1.8$~\cite{supple}. $\Sigma_s$ is the real part of self-energy. Therefore the correction from $\Sigma_s$ to the phonon dispersion can be safely neglected.

Experimentally, the circular phonon dichroism can be detected by the pulse-echo technique based on the different absorption rates between left- and right-handed circular phonons~\cite{truell1969}. Alternatively, we can use the Raman spectroscopy analysis of phonon polarization~\cite{luthi2004}. Our studies can be further generalized to other situations, such as discussing the role of disorder, superconducting states and nonlinear effect in phonon dichroism. These will be the subject of future work.

$\textit{Acknowledgments}$. This work is supported by the National Natural Science Foundation of China (NSFC, Grant No. 11904062), the Starting Research Fund from Guangzhou University (Grant No. RQ2020076) and Guangzhou Basic Research Program, jointed funded by Guangzhou University (Grant No. 202201020186).


\begin{thebibliography}{61}%
\makeatletter
\providecommand \@ifxundefined [1]{%
 \@ifx{#1\undefined}
}%
\providecommand \@ifnum [1]{%
 \ifnum #1\expandafter \@firstoftwo
 \else \expandafter \@secondoftwo
 \fi
}%
\providecommand \@ifx [1]{%
 \ifx #1\expandafter \@firstoftwo
 \else \expandafter \@secondoftwo
 \fi
}%
\providecommand \natexlab [1]{#1}%
\providecommand \enquote  [1]{``#1''}%
\providecommand \bibnamefont  [1]{#1}%
\providecommand \bibfnamefont [1]{#1}%
\providecommand \citenamefont [1]{#1}%
\providecommand \href@noop [0]{\@secondoftwo}%
\providecommand \href [0]{\begingroup \@sanitize@url \@href}%
\providecommand \@href[1]{\@@startlink{#1}\@@href}%
\providecommand \@@href[1]{\endgroup#1\@@endlink}%
\providecommand \@sanitize@url [0]{\catcode `\\12\catcode `\$12\catcode
  `\&12\catcode `\#12\catcode `\^12\catcode `\_12\catcode `\%12\relax}%
\providecommand \@@startlink[1]{}%
\providecommand \@@endlink[0]{}%
\providecommand \url  [0]{\begingroup\@sanitize@url \@url }%
\providecommand \@url [1]{\endgroup\@href {#1}{\urlprefix }}%
\providecommand \urlprefix  [0]{URL }%
\providecommand \Eprint [0]{\href }%
\providecommand \doibase [0]{http://dx.doi.org/}%
\providecommand \selectlanguage [0]{\@gobble}%
\providecommand \bibinfo  [0]{\@secondoftwo}%
\providecommand \bibfield  [0]{\@secondoftwo}%
\providecommand \translation [1]{[#1]}%
\providecommand \BibitemOpen [0]{}%
\providecommand \bibitemStop [0]{}%
\providecommand \bibitemNoStop [0]{.\EOS\space}%
\providecommand \EOS [0]{\spacefactor3000\relax}%
\providecommand \BibitemShut  [1]{\csname bibitem#1\endcsname}%
\let\auto@bib@innerbib\@empty
\bibitem [{\citenamefont {Castro~Neto}\ \emph {et~al.}(2009)\citenamefont
  {Castro~Neto}, \citenamefont {Guinea}, \citenamefont {Peres}, \citenamefont
  {Novoselov},\ and\ \citenamefont {Geim}}]{neto2009}%
  \BibitemOpen
  \bibfield  {author} {\bibinfo {author} {\bibfnamefont {A.~H.}\ \bibnamefont
  {Castro~Neto}}, \bibinfo {author} {\bibfnamefont {F.}~\bibnamefont {Guinea}},
  \bibinfo {author} {\bibfnamefont {N.~M.~R.}\ \bibnamefont {Peres}}, \bibinfo
  {author} {\bibfnamefont {K.~S.}\ \bibnamefont {Novoselov}}, \ and\ \bibinfo
  {author} {\bibfnamefont {A.~K.}\ \bibnamefont {Geim}},\ }\bibfield  {title} {\bibinfo {title} {The electronic properties of graphene}}, \href {\doibase
  10.1103/RevModPhys.81.109} {\bibfield  {journal} {\bibinfo  {journal} {Rev.
  Mod. Phys.}\ }\textbf {\bibinfo {volume} {81}},\ \bibinfo {pages} {109}
  (\bibinfo {year} {2009})}\BibitemShut {NoStop}%
\bibitem [{\citenamefont {Xiao}\ \emph {et~al.}(2012)\citenamefont {Xiao},
  \citenamefont {Liu}, \citenamefont {Feng}, \citenamefont {Xu},\ and\
  \citenamefont {Yao}}]{xiao2012}%
  \BibitemOpen
  \bibfield  {author} {\bibinfo {author} {\bibfnamefont {D.}~\bibnamefont
  {Xiao}}, \bibinfo {author} {\bibfnamefont {G.-B.}\ \bibnamefont {Liu}},
  \bibinfo {author} {\bibfnamefont {W.}~\bibnamefont {Feng}}, \bibinfo {author}
  {\bibfnamefont {X.}~\bibnamefont {Xu}}, \ and\ \bibinfo {author}
  {\bibfnamefont {W.}~\bibnamefont {Yao}},\ }\bibfield  {title} {\bibinfo {title} {Coupled Spin and Valley Physics in Monolayers of ${\mathrm{MoS}}_{2}$ and Other Group-VI Dichalcogenides}}, \href {\doibase
  10.1103/PhysRevLett.108.196802} {\bibfield  {journal} {\bibinfo  {journal}
  {Phys. Rev. Lett.}\ }\textbf {\bibinfo {volume} {108}},\ \bibinfo {pages}
  {196802} (\bibinfo {year} {2012})}\BibitemShut {NoStop}%
\bibitem [{\citenamefont {Berry}(1984)}]{berry1984}%
  \BibitemOpen
  \bibfield  {author} {\bibinfo {author} {\bibfnamefont {M.~V.}\ \bibnamefont
  {Berry}},\ }\bibfield  {title} {\bibinfo {title} {Quantal phase factors accompanying adiabatic changes}}, \href {\doibase 10.1098/rspa.1984.0023} {\bibfield  {journal}
  {\bibinfo  {journal} {Proc. R. Soc. A}\ }\textbf {\bibinfo {volume} {392}},\
  \bibinfo {pages} {45} (\bibinfo {year} {1984})}\BibitemShut {NoStop}%
\bibitem [{\citenamefont {Haldane}(1988)}]{haldane1988}%
  \BibitemOpen
  \bibfield  {author} {\bibinfo {author} {\bibfnamefont {F.~D.~M.}\
  \bibnamefont {Haldane}},\ }\bibfield  {title} {\bibinfo {title} {Model for a Quantum Hall Effect without Landau Levels: Condensed-Matter Realization of the "Parity Anomaly"}}, \href {\doibase 10.1103/PhysRevLett.61.2015}
  {\bibfield  {journal} {\bibinfo  {journal} {Phys. Rev. Lett.}\ }\textbf
  {\bibinfo {volume} {61}},\ \bibinfo {pages} {2015} (\bibinfo {year}
  {1988})}\BibitemShut {NoStop}%
\bibitem [{\citenamefont {Xiao}\ \emph {et~al.}(2010)\citenamefont {Xiao},
  \citenamefont {Chang},\ and\ \citenamefont {Niu}}]{xiao2010}%
  \BibitemOpen
  \bibfield  {author} {\bibinfo {author} {\bibfnamefont {D.}~\bibnamefont
  {Xiao}}, \bibinfo {author} {\bibfnamefont {M.-C.}\ \bibnamefont {Chang}}, \
  and\ \bibinfo {author} {\bibfnamefont {Q.}~\bibnamefont {Niu}},\ }\bibfield  {title} {\bibinfo {title} {Berry phase effects on electronic properties}}, \href
  {\doibase 10.1103/RevModPhys.82.1959} {\bibfield  {journal} {\bibinfo
  {journal} {Rev. Mod. Phys.}\ }\textbf {\bibinfo {volume} {82}},\ \bibinfo
  {pages} {1959} (\bibinfo {year} {2010})}\BibitemShut {NoStop}%
\bibitem [{\citenamefont {Xiao}\ \emph {et~al.}(2007)\citenamefont {Xiao},
  \citenamefont {Yao},\ and\ \citenamefont {Niu}}]{xiao2007}%
  \BibitemOpen
  \bibfield  {author} {\bibinfo {author} {\bibfnamefont {D.}~\bibnamefont
  {Xiao}}, \bibinfo {author} {\bibfnamefont {W.}~\bibnamefont {Yao}}, \ and\
  \bibinfo {author} {\bibfnamefont {Q.}~\bibnamefont {Niu}},\ }\bibfield  {title} {\bibinfo {title} {Valley-Contrasting Physics in Graphene: Magnetic Moment and Topological Transport}}, \href {\doibase
  10.1103/PhysRevLett.99.236809} {\bibfield  {journal} {\bibinfo  {journal}
  {Phys. Rev. Lett.}\ }\textbf {\bibinfo {volume} {99}},\ \bibinfo {pages}
  {236809} (\bibinfo {year} {2007})}\BibitemShut {NoStop}%
\bibitem [{\citenamefont {Yao}\ \emph {et~al.}(2008)\citenamefont {Yao},
  \citenamefont {Xiao},\ and\ \citenamefont {Niu}}]{yao2008}%
  \BibitemOpen
  \bibfield  {author} {\bibinfo {author} {\bibfnamefont {W.}~\bibnamefont
  {Yao}}, \bibinfo {author} {\bibfnamefont {D.}~\bibnamefont {Xiao}}, \ and\
  \bibinfo {author} {\bibfnamefont {Q.}~\bibnamefont {Niu}},\ }\bibfield  {title} {\bibinfo {title} {Valley-dependent optoelectronics from inversion symmetry breaking}}, \href {\doibase
  10.1103/PhysRevB.77.235406} {\bibfield  {journal} {\bibinfo  {journal} {Phys.
  Rev. B}\ }\textbf {\bibinfo {volume} {77}},\ \bibinfo {pages} {235406}
  (\bibinfo {year} {2008})}\BibitemShut {NoStop}%
\bibitem [{\citenamefont {Barkeshli}\ \emph {et~al.}(2012)\citenamefont
  {Barkeshli}, \citenamefont {Chung},\ and\ \citenamefont
  {Qi}}]{barkeshli2012}%
  \BibitemOpen
  \bibfield  {author} {\bibinfo {author} {\bibfnamefont {M.}~\bibnamefont
  {Barkeshli}}, \bibinfo {author} {\bibfnamefont {S.~B.}\ \bibnamefont
  {Chung}}, \ and\ \bibinfo {author} {\bibfnamefont {X.-L.}\ \bibnamefont
  {Qi}},\ }\bibfield  {title} {\bibinfo {title} {Dissipationless phonon Hall viscosity}}, \href {\doibase 10.1103/PhysRevB.85.245107} {\bibfield  {journal}
  {\bibinfo  {journal} {Phys. Rev. B}\ }\textbf {\bibinfo {volume} {85}},\
  \bibinfo {pages} {245107} (\bibinfo {year} {2012})}\BibitemShut {NoStop}%
\bibitem [{\citenamefont {Rinkel}\ \emph {et~al.}(2017)\citenamefont {Rinkel},
  \citenamefont {Lopes},\ and\ \citenamefont {Garate}}]{rinkel2017}%
  \BibitemOpen
  \bibfield  {author} {\bibinfo {author} {\bibfnamefont {P.}~\bibnamefont
  {Rinkel}}, \bibinfo {author} {\bibfnamefont {P.~L.~S.}\ \bibnamefont
  {Lopes}}, \ and\ \bibinfo {author} {\bibfnamefont {I.}~\bibnamefont
  {Garate}},\ }\bibfield  {title} {\bibinfo {title} {Signatures of the Chiral Anomaly in Phonon Dynamics}}, \href {\doibase 10.1103/PhysRevLett.119.107401} {\bibfield
  {journal} {\bibinfo  {journal} {Phys. Rev. Lett.}\ }\textbf {\bibinfo
  {volume} {119}},\ \bibinfo {pages} {107401} (\bibinfo {year}
  {2017})}\BibitemShut {NoStop}%
\bibitem [{\citenamefont {Ren}\ \emph {et~al.}(2021)\citenamefont {Ren},
  \citenamefont {Xiao}, \citenamefont {Saparov},\ and\ \citenamefont
  {Niu}}]{ren2021}%
  \BibitemOpen
  \bibfield  {author} {\bibinfo {author} {\bibfnamefont {Y.}~\bibnamefont
  {Ren}}, \bibinfo {author} {\bibfnamefont {C.}~\bibnamefont {Xiao}}, \bibinfo
  {author} {\bibfnamefont {D.}~\bibnamefont {Saparov}}, \ and\ \bibinfo
  {author} {\bibfnamefont {Q.}~\bibnamefont {Niu}},\ }\bibfield  {title} {\bibinfo {title} {Phonon Magnetic Moment from Electronic Topological Magnetization}}, \href {\doibase
  10.1103/PhysRevLett.127.186403} {\bibfield  {journal} {\bibinfo  {journal}
  {Phys. Rev. Lett.}\ }\textbf {\bibinfo {volume} {127}},\ \bibinfo {pages}
  {186403} (\bibinfo {year} {2021})}\BibitemShut {NoStop}%
\bibitem [{\citenamefont {Spivak}\ and\ \citenamefont
  {Andreev}(2016)}]{spivak2016}%
  \BibitemOpen
  \bibfield  {author} {\bibinfo {author} {\bibfnamefont {B.~Z.}\ \bibnamefont
  {Spivak}}\ and\ \bibinfo {author} {\bibfnamefont {A.~V.}\ \bibnamefont
  {Andreev}},\ }\bibfield  {title} {\bibinfo {title} {Magnetotransport phenomena related to the chiral anomaly in Weyl semimetals}}, \href {\doibase 10.1103/PhysRevB.93.085107} {\bibfield
  {journal} {\bibinfo  {journal} {Phys. Rev. B}\ }\textbf {\bibinfo {volume}
  {93}},\ \bibinfo {pages} {085107} (\bibinfo {year} {2016})}\BibitemShut
  {NoStop}%
\bibitem [{\citenamefont {Sengupta}\ \emph {et~al.}(2020)\citenamefont
  {Sengupta}, \citenamefont {Lhachemi},\ and\ \citenamefont
  {Garate}}]{sengupta2020}%
  \BibitemOpen
  \bibfield  {author} {\bibinfo {author} {\bibfnamefont {S.}~\bibnamefont
  {Sengupta}}, \bibinfo {author} {\bibfnamefont {M.~N.~Y.}\ \bibnamefont
  {Lhachemi}}, \ and\ \bibinfo {author} {\bibfnamefont {I.}~\bibnamefont
  {Garate}},\ }\bibfield  {title} {\bibinfo {title} {Phonon Magnetochiral Effect of Band-Geometric Origin in Weyl Semimetals}}, \href {\doibase 10.1103/PhysRevLett.125.146402} {\bibfield
  {journal} {\bibinfo  {journal} {Phys. Rev. Lett.}\ }\textbf {\bibinfo
  {volume} {125}},\ \bibinfo {pages} {146402} (\bibinfo {year}
  {2020})}\BibitemShut {NoStop}%
\bibitem [{\citenamefont {Hu}\ \emph {et~al.}(2021)\citenamefont {Hu},
  \citenamefont {Yu}, \citenamefont {Garate},\ and\ \citenamefont
  {Liu}}]{hu2021}%
  \BibitemOpen
  \bibfield  {author} {\bibinfo {author} {\bibfnamefont {L.-H.}\ \bibnamefont
  {Hu}}, \bibinfo {author} {\bibfnamefont {J.}~\bibnamefont {Yu}}, \bibinfo
  {author} {\bibfnamefont {I.}~\bibnamefont {Garate}}, \ and\ \bibinfo {author}
  {\bibfnamefont {C.-X.}\ \bibnamefont {Liu}},\ }\bibfield  {title} {\bibinfo {title} {Phonon Helicity Induced by Electronic Berry Curvature in Dirac Materials}}, \href {\doibase
  10.1103/PhysRevLett.127.125901} {\bibfield  {journal} {\bibinfo  {journal}
  {Phys. Rev. Lett.}\ }\textbf {\bibinfo {volume} {127}},\ \bibinfo {pages}
  {125901} (\bibinfo {year} {2021})}\BibitemShut {NoStop}%
\bibitem [{\citenamefont {Li}\ \emph {et~al.}(2012)\citenamefont {Li},
  \citenamefont {Ren}, \citenamefont {Wang}, \citenamefont {Zhang},
  \citenamefont {H\"anggi},\ and\ \citenamefont {Li}}]{li2012}%
  \BibitemOpen
  \bibfield  {author} {\bibinfo {author} {\bibfnamefont {N.}~\bibnamefont
  {Li}}, \bibinfo {author} {\bibfnamefont {J.}~\bibnamefont {Ren}}, \bibinfo
  {author} {\bibfnamefont {L.}~\bibnamefont {Wang}}, \bibinfo {author}
  {\bibfnamefont {G.}~\bibnamefont {Zhang}}, \bibinfo {author} {\bibfnamefont
  {P.}~\bibnamefont {H\"anggi}}, \ and\ \bibinfo {author} {\bibfnamefont
  {B.}~\bibnamefont {Li}},\ }\bibfield  {title} {\bibinfo {title} {Colloquium: Phononics: Manipulating heat flow with electronic analogs and beyond}}, \href {\doibase 10.1103/RevModPhys.84.1045}
  {\bibfield  {journal} {\bibinfo  {journal} {Rev. Mod. Phys.}\ }\textbf
  {\bibinfo {volume} {84}},\ \bibinfo {pages} {1045} (\bibinfo {year}
  {2012})}\BibitemShut {NoStop}%
\bibitem [{\citenamefont {Heil}\ \emph {et~al.}(1982)\citenamefont {Heil},
  \citenamefont {L\"uthi},\ and\ \citenamefont {Thalmeier}}]{heil1982}%
  \BibitemOpen
  \bibfield  {author} {\bibinfo {author} {\bibfnamefont {J.}~\bibnamefont
  {Heil}}, \bibinfo {author} {\bibfnamefont {B.}~\bibnamefont {L\"uthi}}, \
  and\ \bibinfo {author} {\bibfnamefont {P.}~\bibnamefont {Thalmeier}},\ }\bibfield  {title} {\bibinfo {title} {Nonreciprocal surface-acoustic-wave propagation in aluminum}}, \href
  {\doibase 10.1103/PhysRevB.25.6515} {\bibfield  {journal} {\bibinfo
  {journal} {Phys. Rev. B}\ }\textbf {\bibinfo {volume} {25}},\ \bibinfo
  {pages} {6515} (\bibinfo {year} {1982})}\BibitemShut {NoStop}%
\bibitem [{\citenamefont {Liu}\ \emph {et~al.}(2020)\citenamefont {Liu},
  \citenamefont {Guo}, \citenamefont {Hu}, \citenamefont {Shi}, \citenamefont
  {Li}, \citenamefont {Zhang}, \citenamefont {Chen}, \citenamefont {Zhang},
  \citenamefont {Zhou}, \citenamefont {Lu}, \citenamefont {Lin}, \citenamefont
  {Liu}, \citenamefont {Cheng}, \citenamefont {Liu}, \citenamefont {Xie},
  \citenamefont {Bi}, \citenamefont {Tan}, \citenamefont {Deng}, \citenamefont
  {Qiu},\ and\ \citenamefont {Peng}}]{liu2020}%
  \BibitemOpen
  \bibfield  {author} {\bibinfo {author} {\bibfnamefont {Z.}~\bibnamefont
  {Liu}}, \bibinfo {author} {\bibfnamefont {K.}~\bibnamefont {Guo}}, \bibinfo
  {author} {\bibfnamefont {G.}~\bibnamefont {Hu}}, \bibinfo {author}
  {\bibfnamefont {Z.}~\bibnamefont {Shi}}, \bibinfo {author} {\bibfnamefont
  {Y.}~\bibnamefont {Li}}, \bibinfo {author} {\bibfnamefont {L.}~\bibnamefont
  {Zhang}}, \bibinfo {author} {\bibfnamefont {H.}~\bibnamefont {Chen}},
  \bibinfo {author} {\bibfnamefont {L.}~\bibnamefont {Zhang}}, \bibinfo
  {author} {\bibfnamefont {P.}~\bibnamefont {Zhou}}, \bibinfo {author}
  {\bibfnamefont {H.}~\bibnamefont {Lu}}, \bibinfo {author} {\bibfnamefont
  {M.-L.}\ \bibnamefont {Lin}}, \bibinfo {author} {\bibfnamefont
  {S.}~\bibnamefont {Liu}}, \bibinfo {author} {\bibfnamefont {Y.}~\bibnamefont
  {Cheng}}, \bibinfo {author} {\bibfnamefont {X.~L.}\ \bibnamefont {Liu}},
  \bibinfo {author} {\bibfnamefont {J.}~\bibnamefont {Xie}}, \bibinfo {author}
  {\bibfnamefont {L.}~\bibnamefont {Bi}}, \bibinfo {author} {\bibfnamefont
  {P.-H.}\ \bibnamefont {Tan}}, \bibinfo {author} {\bibfnamefont
  {L.}~\bibnamefont {Deng}}, \bibinfo {author} {\bibfnamefont {C.-W.}\
  \bibnamefont {Qiu}}, \ and\ \bibinfo {author} {\bibfnamefont
  {B.}~\bibnamefont {Peng}},\ }\bibfield  {title} {\bibinfo {title} {Observation of nonreciprocal magnetophonon effect in nonencapsulated few-layered CrI$_3$}}, \href {\doibase 10.1126/sciadv.abc7628}
  {\bibfield  {journal} {\bibinfo  {journal} {Sci. Adv.}\ }\textbf {\bibinfo
  {volume} {6}},\ \bibinfo {pages} {eabc7628} (\bibinfo {year}
  {2020})}\BibitemShut {NoStop}%
\bibitem [{\citenamefont {Emtage}(1976)}]{emtage1976}%
  \BibitemOpen
  \bibfield  {author} {\bibinfo {author} {\bibfnamefont {P.~R.}\ \bibnamefont
  {Emtage}},\ }\bibfield  {title} {\bibinfo {title} {Nonreciprocal attenuation of magnetoelastic surface waves}}, \href {\doibase 10.1103/PhysRevB.13.3063} {\bibfield  {journal}
  {\bibinfo  {journal} {Phys. Rev. B}\ }\textbf {\bibinfo {volume} {13}},\
  \bibinfo {pages} {3063} (\bibinfo {year} {1976})}\BibitemShut {NoStop}%
\bibitem [{\citenamefont {Sasaki}\ \emph {et~al.}(2017)\citenamefont {Sasaki},
  \citenamefont {Nii}, \citenamefont {Iguchi},\ and\ \citenamefont
  {Onose}}]{sasaki2017}%
  \BibitemOpen
  \bibfield  {author} {\bibinfo {author} {\bibfnamefont {R.}~\bibnamefont
  {Sasaki}}, \bibinfo {author} {\bibfnamefont {Y.}~\bibnamefont {Nii}},
  \bibinfo {author} {\bibfnamefont {Y.}~\bibnamefont {Iguchi}}, \ and\ \bibinfo
  {author} {\bibfnamefont {Y.}~\bibnamefont {Onose}},\ }\bibfield  {title} {\bibinfo {title} {Nonreciprocal propagation of surface acoustic wave in ${\mathrm{Ni}\text{/}\mathrm{LiNbO}}_{3}$}}, \href {\doibase
  10.1103/PhysRevB.95.020407} {\bibfield  {journal} {\bibinfo  {journal} {Phys.
  Rev. B}\ }\textbf {\bibinfo {volume} {95}},\ \bibinfo {pages} {020407(R)}
  (\bibinfo {year} {2017})}\BibitemShut {NoStop}%
\bibitem [{\citenamefont {Nomura}\ \emph {et~al.}(2019)\citenamefont {Nomura},
  \citenamefont {Zhang}, \citenamefont {Zherlitsyn}, \citenamefont {Wosnitza},
  \citenamefont {Tokura}, \citenamefont {Nagaosa},\ and\ \citenamefont
  {Seki}}]{nomura2019}%
  \BibitemOpen
  \bibfield  {author} {\bibinfo {author} {\bibfnamefont {T.}~\bibnamefont
  {Nomura}}, \bibinfo {author} {\bibfnamefont {X.-X.}\ \bibnamefont {Zhang}},
  \bibinfo {author} {\bibfnamefont {S.}~\bibnamefont {Zherlitsyn}}, \bibinfo
  {author} {\bibfnamefont {J.}~\bibnamefont {Wosnitza}}, \bibinfo {author}
  {\bibfnamefont {Y.}~\bibnamefont {Tokura}}, \bibinfo {author} {\bibfnamefont
  {N.}~\bibnamefont {Nagaosa}}, \ and\ \bibinfo {author} {\bibfnamefont
  {S.}~\bibnamefont {Seki}},\ }\bibfield  {title} {\bibinfo {title} {Phonon Magnetochiral Effect}}, \href {\doibase 10.1103/PhysRevLett.122.145901}
  {\bibfield  {journal} {\bibinfo  {journal} {Phys. Rev. Lett.}\ }\textbf
  {\bibinfo {volume} {122}},\ \bibinfo {pages} {145901} (\bibinfo {year}
  {2019})}\BibitemShut {NoStop}%
\bibitem [{\citenamefont {Xu}\ \emph {et~al.}(2020)\citenamefont {Xu},
  \citenamefont {Yamamoto}, \citenamefont {Puebla}, \citenamefont {Baumgaertl},
  \citenamefont {Rana}, \citenamefont {Miura}, \citenamefont {Takahashi},
  \citenamefont {D.}, \citenamefont {Maekawa},\ and\ \citenamefont
  {Otani}}]{xu2020}%
  \BibitemOpen
  \bibfield  {author} {\bibinfo {author} {\bibfnamefont {M.}~\bibnamefont
  {Xu}}, \bibinfo {author} {\bibfnamefont {K.}~\bibnamefont {Yamamoto}},
  \bibinfo {author} {\bibfnamefont {J.}~\bibnamefont {Puebla}}, \bibinfo
  {author} {\bibfnamefont {K.}~\bibnamefont {Baumgaertl}}, \bibinfo {author}
  {\bibfnamefont {B.}~\bibnamefont {Rana}}, \bibinfo {author} {\bibfnamefont
  {K.}~\bibnamefont {Miura}}, \bibinfo {author} {\bibfnamefont
  {H.}~\bibnamefont {Takahashi}}, \bibinfo {author} {\bibfnamefont
  {G.}~\bibnamefont {D.}}, \bibinfo {author} {\bibfnamefont {S.}~\bibnamefont
  {Maekawa}}, \ and\ \bibinfo {author} {\bibfnamefont {Y.}~\bibnamefont
  {Otani}},\ }\bibfield  {title} {\bibinfo {title} {Nonreciprocal surface acoustic wave propagation via magneto-rotation coupling}}, \href {\doibase 10.1126/sciadv.abb1724} {\bibfield  {journal}
  {\bibinfo  {journal} {Sci. Adv.}\ }\textbf {\bibinfo {volume} {6}},\ \bibinfo
  {pages} {eabb1724} (\bibinfo {year} {2020})}\BibitemShut {NoStop}%
\bibitem [{\citenamefont {Shan}\ \emph {et~al.}(2020)\citenamefont {Shan},
  \citenamefont {Bas}, \citenamefont {I.}, \citenamefont {A.}, \citenamefont
  {Sun},\ and\ \citenamefont {Page}}]{shanp2020}%
  \BibitemOpen
  \bibfield  {author} {\bibinfo {author} {\bibfnamefont {P.~J.}\ \bibnamefont
  {Shan}}, \bibinfo {author} {\bibfnamefont {D.~A.}\ \bibnamefont {Bas}},
  \bibinfo {author} {\bibfnamefont {L.}~\bibnamefont {I.}}, \bibinfo {author}
  {\bibfnamefont {M.}~\bibnamefont {A.}}, \bibinfo {author} {\bibfnamefont
  {N.~X.}\ \bibnamefont {Sun}}, \ and\ \bibinfo {author} {\bibfnamefont
  {M.~R.}\ \bibnamefont {Page}},\ }\bibfield  {title} {\bibinfo {title} {Giant nonreciprocity of surface acoustic waves enabled by the magnetoelastic interaction}}, \href {\doibase 10.1126/sciadv.abc5648}
  {\bibfield  {journal} {\bibinfo  {journal} {Sci. Adv.}\ }\textbf {\bibinfo
  {volume} {6}},\ \bibinfo {pages} {eabc5648} (\bibinfo {year}
  {2020})}\BibitemShut {NoStop}%
\bibitem [{\citenamefont {Hern\'andez-M\'{\i}nguez}\ \emph
  {et~al.}(2020)\citenamefont {Hern\'andez-M\'{\i}nguez}, \citenamefont
  {Maci\`a}, \citenamefont {Hern\`andez}, \citenamefont {Herfort},\ and\
  \citenamefont {Santos}}]{hernandez2020}%
  \BibitemOpen
  \bibfield  {author} {\bibinfo {author} {\bibfnamefont {A.}~\bibnamefont
  {Hern\'andez-M\'{\i}nguez}}, \bibinfo {author} {\bibfnamefont
  {F.}~\bibnamefont {Maci\`a}}, \bibinfo {author} {\bibfnamefont {J.~M.}\
  \bibnamefont {Hern\`andez}}, \bibinfo {author} {\bibfnamefont
  {J.}~\bibnamefont {Herfort}}, \ and\ \bibinfo {author} {\bibfnamefont
  {P.~V.}\ \bibnamefont {Santos}},\ }\bibfield  {title} {\bibinfo {title} {Large Nonreciprocal Propagation of Surface Acoustic Waves in Epitaxial Ferromagnetic/Semiconductor Hybrid Structures}}, \href {\doibase
  10.1103/PhysRevApplied.13.044018} {\bibfield  {journal} {\bibinfo  {journal}
  {Phys. Rev. Applied}\ }\textbf {\bibinfo {volume} {13}},\ \bibinfo {pages}
  {044018} (\bibinfo {year} {2020})}\BibitemShut {NoStop}%
\bibitem [{\citenamefont {Tateno}\ and\ \citenamefont
  {Nozaki}(2020)}]{tateno2020}%
  \BibitemOpen
  \bibfield  {author} {\bibinfo {author} {\bibfnamefont {S.}~\bibnamefont
  {Tateno}}\ and\ \bibinfo {author} {\bibfnamefont {Y.}~\bibnamefont
  {Nozaki}},\ }\bibfield  {title} {\bibinfo {title} {Highly Nonreciprocal Spin Waves Excited by Magnetoelastic Coupling in a $\mathrm{Ni}$/$\mathrm{Si}$ Bilayer}}, \href {\doibase 10.1103/PhysRevApplied.13.034074} {\bibfield
  {journal} {\bibinfo  {journal} {Phys. Rev. Applied}\ }\textbf {\bibinfo
  {volume} {13}},\ \bibinfo {pages} {034074} (\bibinfo {year}
  {2020})}\BibitemShut {NoStop}%
\bibitem [{\citenamefont {Liang}\ \emph {et~al.}(2010)\citenamefont {Liang},
  \citenamefont {Guo}, \citenamefont {Tu}, \citenamefont {Zhang},\ and\
  \citenamefont {Cheng}}]{liang2010}%
  \BibitemOpen
  \bibfield  {author} {\bibinfo {author} {\bibfnamefont {B.}~\bibnamefont
  {Liang}}, \bibinfo {author} {\bibfnamefont {X.~S.}\ \bibnamefont {Guo}},
  \bibinfo {author} {\bibfnamefont {J.}~\bibnamefont {Tu}}, \bibinfo {author}
  {\bibfnamefont {D.}~\bibnamefont {Zhang}}, \ and\ \bibinfo {author}
  {\bibfnamefont {J.~C.}\ \bibnamefont {Cheng}},\ }\bibfield  {title} {\bibinfo {title} {An acoustic rectifier}}, \href {\doibase
  10.1038/nmat2881} {\bibfield  {journal} {\bibinfo  {journal} {Nature
  Materials}\ }\textbf {\bibinfo {volume} {9}},\ \bibinfo {pages} {989}
  (\bibinfo {year} {2010})}\BibitemShut {NoStop}%
\bibitem [{\citenamefont {Xu}\ \emph {et~al.}(2019)\citenamefont {Xu},
  \citenamefont {Jiang}, \citenamefont {Clerk},\ and\ \citenamefont
  {Harris}}]{xu2019}%
  \BibitemOpen
  \bibfield  {author} {\bibinfo {author} {\bibfnamefont {H.}~\bibnamefont
  {Xu}}, \bibinfo {author} {\bibfnamefont {L.}~\bibnamefont {Jiang}}, \bibinfo
  {author} {\bibfnamefont {A.~A.}\ \bibnamefont {Clerk}}, \ and\ \bibinfo
  {author} {\bibfnamefont {J.~G.~E.}\ \bibnamefont {Harris}},\ }\bibfield  {title} {\bibinfo {title} {Nonreciprocal control and cooling of phonon modes in an optomechanical system}}, \href {\doibase
  10.1038/s41586-019-1061-2} {\bibfield  {journal} {\bibinfo  {journal}
  {Nature}\ }\textbf {\bibinfo {volume} {568}},\ \bibinfo {pages} {65}
  (\bibinfo {year} {2019})}\BibitemShut {NoStop}%
\bibitem [{\citenamefont {Ge}\ and\ \citenamefont {Liu}(2013)}]{ge2013}%
  \BibitemOpen
  \bibfield  {author} {\bibinfo {author} {\bibfnamefont {Y.}~\bibnamefont
  {Ge}}\ and\ \bibinfo {author} {\bibfnamefont {A.~Y.}\ \bibnamefont {Liu}},\
  }\bibfield  {title} {\bibinfo {title} {Phonon-mediated superconductivity in electron-doped single-layer MoS${}_{2}$: A first-principles prediction}}, \href {\doibase 10.1103/PhysRevB.87.241408} {\bibfield  {journal} {\bibinfo
  {journal} {Phys. Rev. B}\ }\textbf {\bibinfo {volume} {87}},\ \bibinfo
  {pages} {241408(R)} (\bibinfo {year} {2013})}\BibitemShut {NoStop}%
\bibitem [{\citenamefont {Navarro-Moratalla}\ \emph {et~al.}(2016)\citenamefont
  {Navarro-Moratalla}, \citenamefont {Island}, \citenamefont
  {Ma{\~n}as-Valero}, \citenamefont {Pinilla-Cienfuegos}, \citenamefont
  {Castellanos-Gomez}, \citenamefont {Quereda}, \citenamefont
  {Rubio-Bollinger}, \citenamefont {Chirolli}, \citenamefont
  {Silva-Guill{\'e}n}, \citenamefont {Agra{\"\i}t}, \citenamefont {Steele},
  \citenamefont {Guinea}, \citenamefont {van~der Zant},\ and\ \citenamefont
  {Coronado}}]{navarro2016}%
  \BibitemOpen
  \bibfield  {author} {\bibinfo {author} {\bibfnamefont {E.}~\bibnamefont
  {Navarro-Moratalla}}, \bibinfo {author} {\bibfnamefont {J.~O.}\ \bibnamefont
  {Island}}, \bibinfo {author} {\bibfnamefont {S.}~\bibnamefont
  {Ma{\~n}as-Valero}}, \bibinfo {author} {\bibfnamefont {E.}~\bibnamefont
  {Pinilla-Cienfuegos}}, \bibinfo {author} {\bibfnamefont {A.}~\bibnamefont
  {Castellanos-Gomez}}, \bibinfo {author} {\bibfnamefont {J.}~\bibnamefont
  {Quereda}}, \bibinfo {author} {\bibfnamefont {G.}~\bibnamefont
  {Rubio-Bollinger}}, \bibinfo {author} {\bibfnamefont {L.}~\bibnamefont
  {Chirolli}}, \bibinfo {author} {\bibfnamefont {J.~A.}\ \bibnamefont
  {Silva-Guill{\'e}n}}, \bibinfo {author} {\bibfnamefont {N.}~\bibnamefont
  {Agra{\"\i}t}}, \bibinfo {author} {\bibfnamefont {G.~A.}\ \bibnamefont
  {Steele}}, \bibinfo {author} {\bibfnamefont {F.}~\bibnamefont {Guinea}},
  \bibinfo {author} {\bibfnamefont {H.~S.~J.}\ \bibnamefont {van~der Zant}}, \
  and\ \bibinfo {author} {\bibfnamefont {E.}~\bibnamefont {Coronado}},\ }\bibfield  {title} {\bibinfo {title} {Enhanced superconductivity in atomically thin TaS$_2$}}, \href
  {\doibase 10.1038/ncomms11043} {\bibfield  {journal} {\bibinfo  {journal}
  {Nat. Commun.}\ }\textbf {\bibinfo {volume} {7}},\ \bibinfo {pages} {11043}
  (\bibinfo {year} {2016})}\BibitemShut {NoStop}%
\bibitem [{\citenamefont {Choi}\ and\ \citenamefont {Choi}(2018)}]{choi2018}%
  \BibitemOpen
  \bibfield  {author} {\bibinfo {author} {\bibfnamefont {Y.~W.}\ \bibnamefont
  {Choi}}\ and\ \bibinfo {author} {\bibfnamefont {H.~J.}\ \bibnamefont
  {Choi}},\ }\bibfield  {title} {\bibinfo {title} {Strong electron-phonon coupling, electron-hole asymmetry, and nonadiabaticity in magic-angle twisted bilayer graphene}}, \href {\doibase 10.1103/PhysRevB.98.241412} {\bibfield  {journal}
  {\bibinfo  {journal} {Phys. Rev. B}\ }\textbf {\bibinfo {volume} {98}},\
  \bibinfo {pages} {241412(R)} (\bibinfo {year} {2018})}\BibitemShut {NoStop}%
\bibitem [{\citenamefont {Sohier}\ \emph {et~al.}(2019)\citenamefont {Sohier},
  \citenamefont {Ponomarev}, \citenamefont {Gibertini}, \citenamefont {Berger},
  \citenamefont {Marzari}, \citenamefont {Ubrig},\ and\ \citenamefont
  {Morpurgo}}]{sohier2019}%
  \BibitemOpen
  \bibfield  {author} {\bibinfo {author} {\bibfnamefont {T.}~\bibnamefont
  {Sohier}}, \bibinfo {author} {\bibfnamefont {E.}~\bibnamefont {Ponomarev}},
  \bibinfo {author} {\bibfnamefont {M.}~\bibnamefont {Gibertini}}, \bibinfo
  {author} {\bibfnamefont {H.}~\bibnamefont {Berger}}, \bibinfo {author}
  {\bibfnamefont {N.}~\bibnamefont {Marzari}}, \bibinfo {author} {\bibfnamefont
  {N.}~\bibnamefont {Ubrig}}, \ and\ \bibinfo {author} {\bibfnamefont {A.~F.}\
  \bibnamefont {Morpurgo}},\ }\bibfield  {title} {\bibinfo {title} {Enhanced Electron-Phonon Interaction in Multivalley Materials}}, \href {\doibase 10.1103/PhysRevX.9.031019}
  {\bibfield  {journal} {\bibinfo  {journal} {Phys. Rev. X}\ }\textbf {\bibinfo
  {volume} {9}},\ \bibinfo {pages} {031019} (\bibinfo {year}
  {2019})}\BibitemShut {NoStop}%
\bibitem [{\citenamefont {Han}\ \emph {et~al.}(2021)\citenamefont {Han},
  \citenamefont {Chen}, \citenamefont {Cai}, \citenamefont {Wang},
  \citenamefont {Wang}, \citenamefont {Xin},\ and\ \citenamefont
  {Zhang}}]{han2021}%
  \BibitemOpen
  \bibfield  {author} {\bibinfo {author} {\bibfnamefont {T.~T.}\ \bibnamefont
  {Han}}, \bibinfo {author} {\bibfnamefont {L.}~\bibnamefont {Chen}}, \bibinfo
  {author} {\bibfnamefont {C.}~\bibnamefont {Cai}}, \bibinfo {author}
  {\bibfnamefont {Z.~G.}\ \bibnamefont {Wang}}, \bibinfo {author}
  {\bibfnamefont {Y.~D.}\ \bibnamefont {Wang}}, \bibinfo {author}
  {\bibfnamefont {Z.~M.}\ \bibnamefont {Xin}}, \ and\ \bibinfo {author}
  {\bibfnamefont {Y.}~\bibnamefont {Zhang}},\ }\bibfield  {title} {\bibinfo {title} {Metal-Insulator Transition and Emergent Gapped Phase in the Surface-Doped 2D Semiconductor $2\mathrm{H}\text{\ensuremath{-}}{\mathrm{MoTe}}_{2}$}}, \href {\doibase
  10.1103/PhysRevLett.126.106602} {\bibfield  {journal} {\bibinfo  {journal}
  {Phys. Rev. Lett.}\ }\textbf {\bibinfo {volume} {126}},\ \bibinfo {pages}
  {106602} (\bibinfo {year} {2021})}\BibitemShut {NoStop}%
\bibitem [{\citenamefont {Shan}(2022)}]{shan2022}%
  \BibitemOpen
  \bibfield  {author} {\bibinfo {author} {\bibfnamefont {W.-Y.}\ \bibnamefont
  {Shan}},\ }\bibfield  {title} {\bibinfo {title} {Anomalous circular phonon dichroism in transition metal dichalcogenides}}, \href {\doibase 10.1103/PhysRevB.105.L121302} {\bibfield
  {journal} {\bibinfo  {journal} {Phys. Rev. B}\ }\textbf {\bibinfo {volume}
  {105}},\ \bibinfo {pages} {L121302} (\bibinfo {year} {2022})}\BibitemShut
  {NoStop}%
\bibitem [{\citenamefont {Szaller}\ \emph {et~al.}(2013)\citenamefont
  {Szaller}, \citenamefont {Bord\'acs},\ and\ \citenamefont
  {K\'ezsm\'arki}}]{szaller2013}%
  \BibitemOpen
  \bibfield  {author} {\bibinfo {author} {\bibfnamefont {D.}~\bibnamefont
  {Szaller}}, \bibinfo {author} {\bibfnamefont {S.}~\bibnamefont {Bord\'acs}},
  \ and\ \bibinfo {author} {\bibfnamefont {I.}~\bibnamefont {K\'ezsm\'arki}},\
  }\bibfield  {title} {\bibinfo {title} {Symmetry conditions for nonreciprocal light propagation in magnetic crystals}}, \href {\doibase 10.1103/PhysRevB.87.014421} {\bibfield  {journal} {\bibinfo
  {journal} {Phys. Rev. B}\ }\textbf {\bibinfo {volume} {87}},\ \bibinfo
  {pages} {014421} (\bibinfo {year} {2013})}\BibitemShut {NoStop}%
\bibitem [{\citenamefont {Cheong}\ \emph {et~al.}(2018)\citenamefont {Cheong},
  \citenamefont {Talbayev}, \citenamefont {Kiryukhin},\ and\ \citenamefont
  {Saxena}}]{cheong2018}%
  \BibitemOpen
  \bibfield  {author} {\bibinfo {author} {\bibfnamefont {S.-W.}\ \bibnamefont
  {Cheong}}, \bibinfo {author} {\bibfnamefont {D.}~\bibnamefont {Talbayev}},
  \bibinfo {author} {\bibfnamefont {V.}~\bibnamefont {Kiryukhin}}, \ and\
  \bibinfo {author} {\bibfnamefont {A.}~\bibnamefont {Saxena}},\ }\bibfield  {title} {\bibinfo {title} {Broken symmetries, non-reciprocity, and multiferroicity}}, \href
  {\doibase 10.1038/s41535-018-0092-5} {\bibfield  {journal} {\bibinfo
  {journal} {npj Quant. Mater.}\ }\textbf {\bibinfo {volume} {3}},\ \bibinfo
  {pages} {19} (\bibinfo {year} {2018})}\BibitemShut {NoStop}%
\bibitem [{\citenamefont {Tokura}\ and\ \citenamefont
  {Nagaosa}(2018)}]{tokura2018}%
  \BibitemOpen
  \bibfield  {author} {\bibinfo {author} {\bibfnamefont {Y.}~\bibnamefont
  {Tokura}}\ and\ \bibinfo {author} {\bibfnamefont {N.}~\bibnamefont
  {Nagaosa}},\ }\bibfield  {title} {\bibinfo {title} {Nonreciprocal responses from non-centrosymmetric quantum materials}}, \href {\doibase 10.1038/s41467-018-05759-4} {\bibfield
  {journal} {\bibinfo  {journal} {Nat. Commun.}\ }\textbf {\bibinfo {volume}
  {9}},\ \bibinfo {pages} {3740} (\bibinfo {year} {2018})}\BibitemShut
  {NoStop}%
\bibitem [{\citenamefont {Yu}\ \emph {et~al.}(2023)\citenamefont {Yu},
  \citenamefont {Luo},\ and\ \citenamefont {Bauer}}]{yu2023}%
  \BibitemOpen
  \bibfield  {author} {\bibinfo {author} {\bibfnamefont {T.}~\bibnamefont
  {Yu}}, \bibinfo {author} {\bibfnamefont {Z.}~\bibnamefont {Luo}}, \ and\
  \bibinfo {author} {\bibfnamefont {G.~E.}\ \bibnamefont {Bauer}},\ }\bibfield  {title} {\bibinfo {title} {Chirality as generalized spin-orbit interaction in spintronics}}, \href
  {\doibase 10.1016/j.physrep.2023.01.002} {\bibfield  {journal} {\bibinfo
  {journal} {Physics Reports}\ }\textbf {\bibinfo {volume} {1009}},\ \bibinfo
  {pages} {1} (\bibinfo {year} {2023})}\BibitemShut {NoStop}%
\bibitem [{lan()}]{landau1959}%
  \BibitemOpen
  \href@noop {} {}\bibinfo {note} {L. D. Landau and E. M. Lifschitz,
  \emph{Theory of Elasticity}, Pergamon Press, Oxford, 1959.}\BibitemShut
  {Stop}%
\bibitem [{\citenamefont {Suzuura}\ and\ \citenamefont
  {Ando}(2002)}]{suzuura2002}%
  \BibitemOpen
  \bibfield  {author} {\bibinfo {author} {\bibfnamefont {H.}~\bibnamefont
  {Suzuura}}\ and\ \bibinfo {author} {\bibfnamefont {T.}~\bibnamefont {Ando}},\
  }\bibfield  {title} {\bibinfo {title} {Phonons and electron-phonon scattering in carbon nanotubes}}, \href {\doibase 10.1103/PhysRevB.65.235412} {\bibfield  {journal} {\bibinfo
  {journal} {Phys. Rev. B}\ }\textbf {\bibinfo {volume} {65}},\ \bibinfo
  {pages} {235412} (\bibinfo {year} {2002})}\BibitemShut {NoStop}%
\bibitem [{\citenamefont {Liu}\ and\ \citenamefont {Shi}(2017)}]{liu2017}%
  \BibitemOpen
  \bibfield  {author} {\bibinfo {author} {\bibfnamefont {D.}~\bibnamefont
  {Liu}}\ and\ \bibinfo {author} {\bibfnamefont {J.}~\bibnamefont {Shi}},\
  }\bibfield  {title} {\bibinfo {title} {Circular phonon dichroism in Weyl semimetals}}, \href {\doibase 10.1103/PhysRevLett.119.075301} {\bibfield  {journal}
  {\bibinfo  {journal} {Phys. Rev. Lett.}\ }\textbf {\bibinfo {volume} {119}},\
  \bibinfo {pages} {075301} (\bibinfo {year} {2017})}\BibitemShut {NoStop}%
\bibitem [{\citenamefont {Shan}(2020)}]{shan2020}%
  \BibitemOpen
  \bibfield  {author} {\bibinfo {author} {\bibfnamefont {W.-Y.}\ \bibnamefont
  {Shan}},\ }\bibfield  {title} {\bibinfo {title} {Impact of novel electron-phonon coupling mechanisms on valley physics in two-dimensional materials}}, \href {\doibase 10.1103/PhysRevB.102.241301} {\bibfield  {journal}
  {\bibinfo  {journal} {Phys. Rev. B}\ }\textbf {\bibinfo {volume} {102}},\
  \bibinfo {pages} {241301(R)} (\bibinfo {year} {2020})}\BibitemShut {NoStop}%
\bibitem [{sup()}]{supple}%
  \BibitemOpen
  \href@noop {} {}\bibinfo {note} {See Supplementary Material for calculation
  details.}\BibitemShut {Stop}%
\bibitem [{mah()}]{mahan2000}%
  \BibitemOpen
  \href@noop {} {}\bibinfo {note} {G. D. Mahan, \emph{Many-Particle Physics},
  Springer, 2000.}\BibitemShut {Stop}%
\bibitem [{giu()}]{giuliani2005}%
  \BibitemOpen
  \href@noop {} {}\bibinfo {note} {G. Giuliani, and G. Vignale, \emph{Quantum
  Theory of the Electron Liquid}, First Edition (Cambridge University Press,
  2005).}\BibitemShut {Stop}%
\bibitem [{\citenamefont {Bhalla}\ \emph {et~al.}(2022)\citenamefont {Bhalla},
  \citenamefont {Vignale},\ and\ \citenamefont {Rostami}}]{bhalla2022}%
  \BibitemOpen
  \bibfield  {author} {\bibinfo {author} {\bibfnamefont {P.}~\bibnamefont
  {Bhalla}}, \bibinfo {author} {\bibfnamefont {G.}~\bibnamefont {Vignale}}, \
  and\ \bibinfo {author} {\bibfnamefont {H.}~\bibnamefont {Rostami}},\ }\bibfield  {title} {\bibinfo {title} {Pseudogauge field driven acoustoelectric current in two-dimensional hexagonal Dirac materials}}, \href
  {\doibase 10.1103/PhysRevB.105.125407} {\bibfield  {journal} {\bibinfo
  {journal} {Phys. Rev. B}\ }\textbf {\bibinfo {volume} {105}},\ \bibinfo
  {pages} {125407} (\bibinfo {year} {2022})}\BibitemShut {NoStop}%
\bibitem [{\citenamefont {Kane}\ and\ \citenamefont {Mele}(2005)}]{kane2005}%
  \BibitemOpen
  \bibfield  {author} {\bibinfo {author} {\bibfnamefont {C.~L.}\ \bibnamefont
  {Kane}}\ and\ \bibinfo {author} {\bibfnamefont {E.~J.}\ \bibnamefont
  {Mele}},\ }\bibfield  {title} {\bibinfo {title} {Quantum Spin Hall Effect in Graphene}}, \href {\doibase 10.1103/PhysRevLett.95.226801} {\bibfield
  {journal} {\bibinfo  {journal} {Phys. Rev. Lett.}\ }\textbf {\bibinfo
  {volume} {95}},\ \bibinfo {pages} {226801} (\bibinfo {year}
  {2005})}\BibitemShut {NoStop}%
\bibitem [{\citenamefont {Qi}\ \emph {et~al.}(2015)\citenamefont {Qi},
  \citenamefont {Li}, \citenamefont {Niu},\ and\ \citenamefont
  {Feng}}]{qi2015}%
  \BibitemOpen
  \bibfield  {author} {\bibinfo {author} {\bibfnamefont {J.}~\bibnamefont
  {Qi}}, \bibinfo {author} {\bibfnamefont {X.}~\bibnamefont {Li}}, \bibinfo
  {author} {\bibfnamefont {Q.}~\bibnamefont {Niu}}, \ and\ \bibinfo {author}
  {\bibfnamefont {J.}~\bibnamefont {Feng}},\ }\bibfield  {title} {\bibinfo {title} {Giant and tunable valley degeneracy splitting in ${\mathrm{MoTe}}_{2}$}}, \href {\doibase
  10.1103/PhysRevB.92.121403} {\bibfield  {journal} {\bibinfo  {journal} {Phys.
  Rev. B}\ }\textbf {\bibinfo {volume} {92}},\ \bibinfo {pages} {121403(R)}
  (\bibinfo {year} {2015})}\BibitemShut {NoStop}%
\bibitem [{\citenamefont {Norden}\ \emph {et~al.}(2019)\citenamefont {Norden},
  \citenamefont {Zhao}, \citenamefont {Zhang}, \citenamefont {Sabirianov},
  \citenamefont {Petrou},\ and\ \citenamefont {Zeng}}]{norden2019}%
  \BibitemOpen
  \bibfield  {author} {\bibinfo {author} {\bibfnamefont {T.}~\bibnamefont
  {Norden}}, \bibinfo {author} {\bibfnamefont {C.}~\bibnamefont {Zhao}},
  \bibinfo {author} {\bibfnamefont {P.}~\bibnamefont {Zhang}}, \bibinfo
  {author} {\bibfnamefont {R.}~\bibnamefont {Sabirianov}}, \bibinfo {author}
  {\bibfnamefont {A.}~\bibnamefont {Petrou}}, \ and\ \bibinfo {author}
  {\bibfnamefont {H.}~\bibnamefont {Zeng}},\ }\bibfield  {title} {\bibinfo {title} {Giant valley splitting in monolayer WS$_2$ by magnetic proximity effect}}, \href {\doibase
  10.1038/s41467-019-11966-4} {\bibfield  {journal} {\bibinfo  {journal} {Nat.
  Commun.}\ }\textbf {\bibinfo {volume} {10}},\ \bibinfo {pages} {4163}
  (\bibinfo {year} {2019})}\BibitemShut {NoStop}%
\bibitem [{\citenamefont {Cui}\ \emph {et~al.}(2021)\citenamefont {Cui},
  \citenamefont {Zhu}, \citenamefont {Liang}, \citenamefont {Cui},\ and\
  \citenamefont {Yang}}]{cui2021}%
  \BibitemOpen
  \bibfield  {author} {\bibinfo {author} {\bibfnamefont {Q.}~\bibnamefont
  {Cui}}, \bibinfo {author} {\bibfnamefont {Y.}~\bibnamefont {Zhu}}, \bibinfo
  {author} {\bibfnamefont {J.}~\bibnamefont {Liang}}, \bibinfo {author}
  {\bibfnamefont {P.}~\bibnamefont {Cui}}, \ and\ \bibinfo {author}
  {\bibfnamefont {H.}~\bibnamefont {Yang}},\ }\bibfield  {title} {\bibinfo {title} {Spin-valley coupling in a two-dimensional $\mathrm{V}{\mathrm{Si}}_{2}{\mathrm{N}}_{4}$ monolayer}}, \href {\doibase
  10.1103/PhysRevB.103.085421} {\bibfield  {journal} {\bibinfo  {journal}
  {Phys. Rev. B}\ }\textbf {\bibinfo {volume} {103}},\ \bibinfo {pages}
  {085421} (\bibinfo {year} {2021})}\BibitemShut {NoStop}%
\bibitem [{\citenamefont {Chen}\ \emph {et~al.}(2016)\citenamefont {Chen},
  \citenamefont {Luo}, \citenamefont {Xiao}, \citenamefont {Lu}, \citenamefont
  {Zhang}, \citenamefont {Yang}, \citenamefont {Li}, \citenamefont {Pei},
  \citenamefont {Shao}, \citenamefont {Zhang}, \citenamefont {Ling},
  \citenamefont {Xi}, \citenamefont {Song},\ and\ \citenamefont
  {Sun}}]{chen2016}%
  \BibitemOpen
  \bibfield  {author} {\bibinfo {author} {\bibfnamefont {F.~C.}\ \bibnamefont
  {Chen}}, \bibinfo {author} {\bibfnamefont {X.}~\bibnamefont {Luo}}, \bibinfo
  {author} {\bibfnamefont {R.~C.}\ \bibnamefont {Xiao}}, \bibinfo {author}
  {\bibfnamefont {W.~J.}\ \bibnamefont {Lu}}, \bibinfo {author} {\bibfnamefont
  {B.}~\bibnamefont {Zhang}}, \bibinfo {author} {\bibfnamefont {H.~X.}\
  \bibnamefont {Yang}}, \bibinfo {author} {\bibfnamefont {J.~Q.}\ \bibnamefont
  {Li}}, \bibinfo {author} {\bibfnamefont {Q.~L.}\ \bibnamefont {Pei}},
  \bibinfo {author} {\bibfnamefont {D.~F.}\ \bibnamefont {Shao}}, \bibinfo
  {author} {\bibfnamefont {R.~R.}\ \bibnamefont {Zhang}}, \bibinfo {author}
  {\bibfnamefont {L.~S.}\ \bibnamefont {Ling}}, \bibinfo {author}
  {\bibfnamefont {C.~Y.}\ \bibnamefont {Xi}}, \bibinfo {author} {\bibfnamefont
  {W.~H.}\ \bibnamefont {Song}}, \ and\ \bibinfo {author} {\bibfnamefont
  {Y.~P.}\ \bibnamefont {Sun}},\ }\bibfield  {title} {\bibinfo {title} {Superconductivity enhancement in the S-doped Weyl semimetal candidate MoTe2}}, \href {\doibase 10.1063/1.4947433} {\bibfield
   {journal} {\bibinfo  {journal} {Appl. Phys. Lett.}\ }\textbf {\bibinfo
  {volume} {108}},\ \bibinfo {pages} {162601} (\bibinfo {year}
  {2016})}\BibitemShut {NoStop}%
\bibitem [{\citenamefont {Rano}\ \emph {et~al.}(2020)\citenamefont {Rano},
  \citenamefont {Syed},\ and\ \citenamefont {Naqib}}]{rano2020}%
  \BibitemOpen
  \bibfield  {author} {\bibinfo {author} {\bibfnamefont {B.~R.}\ \bibnamefont
  {Rano}}, \bibinfo {author} {\bibfnamefont {I.~M.}\ \bibnamefont {Syed}}, \
  and\ \bibinfo {author} {\bibfnamefont {S.~H.}\ \bibnamefont {Naqib}},\ }\bibfield  {title} {\bibinfo {title} {Elastic, electronic, bonding, and optical properties of WTe$_2$ Weyl semimetal: A comparative investigation with MoTe2 from first principles}}, \href
  {\doibase 10.1016/j.rinp.2020.103639} {\bibfield  {journal} {\bibinfo
  {journal} {Results in Physics}\ }\textbf {\bibinfo {volume} {19}},\ \bibinfo
  {pages} {103639} (\bibinfo {year} {2020})}\BibitemShut {NoStop}%
\bibitem [{\citenamefont {Onsager}(1931)}]{onsager1931}%
  \BibitemOpen
  \bibfield  {author} {\bibinfo {author} {\bibfnamefont {L.}~\bibnamefont
  {Onsager}},\ }\bibfield  {title} {\bibinfo {title} {Reciprocal Relations in Irreversible Processes. I.}}, \href {\doibase 10.1103/PhysRev.37.405} {\bibfield  {journal}
  {\bibinfo  {journal} {Phys. Rev.}\ }\textbf {\bibinfo {volume} {37}},\
  \bibinfo {pages} {405} (\bibinfo {year} {1931})}\BibitemShut {NoStop}%
\bibitem [{\citenamefont {Rikken}\ \emph {et~al.}(2001)\citenamefont {Rikken},
  \citenamefont {F\"olling},\ and\ \citenamefont {Wyder}}]{rikken2001}%
  \BibitemOpen
  \bibfield  {author} {\bibinfo {author} {\bibfnamefont {G.~L. J.~A.}\
  \bibnamefont {Rikken}}, \bibinfo {author} {\bibfnamefont {J.}~\bibnamefont
  {F\"olling}}, \ and\ \bibinfo {author} {\bibfnamefont {P.}~\bibnamefont
  {Wyder}},\ }\bibfield  {title} {\bibinfo {title} {Electrical Magnetochiral Anisotropy}}, \href {\doibase 10.1103/PhysRevLett.87.236602} {\bibfield
  {journal} {\bibinfo  {journal} {Phys. Rev. Lett.}\ }\textbf {\bibinfo
  {volume} {87}},\ \bibinfo {pages} {236602} (\bibinfo {year}
  {2001})}\BibitemShut {NoStop}%
\bibitem [{\citenamefont {Liu}\ \emph {et~al.}(2013)\citenamefont {Liu},
  \citenamefont {Shan}, \citenamefont {Yao}, \citenamefont {Yao},\ and\
  \citenamefont {Xiao}}]{liu2013}%
  \BibitemOpen
  \bibfield  {author} {\bibinfo {author} {\bibfnamefont {G.-B.}\ \bibnamefont
  {Liu}}, \bibinfo {author} {\bibfnamefont {W.-Y.}\ \bibnamefont {Shan}},
  \bibinfo {author} {\bibfnamefont {Y.}~\bibnamefont {Yao}}, \bibinfo {author}
  {\bibfnamefont {W.}~\bibnamefont {Yao}}, \ and\ \bibinfo {author}
  {\bibfnamefont {D.}~\bibnamefont {Xiao}},\ }\bibfield  {title} {\bibinfo {title} {Three-band tight-binding model for monolayers of group-VIB transition metal dichalcogenides}}, \href {\doibase
  10.1103/PhysRevB.88.085433} {\bibfield  {journal} {\bibinfo  {journal} {Phys.
  Rev. B}\ }\textbf {\bibinfo {volume} {88}},\ \bibinfo {pages} {085433}
  (\bibinfo {year} {2013})}\BibitemShut {NoStop}%
\bibitem [{\citenamefont {Korm\'anyos}\ \emph {et~al.}(2013)\citenamefont
  {Korm\'anyos}, \citenamefont {Z\'olyomi}, \citenamefont {Drummond},
  \citenamefont {Rakyta}, \citenamefont {Burkard},\ and\ \citenamefont
  {Fal'ko}}]{kormanyos2013}%
  \BibitemOpen
  \bibfield  {author} {\bibinfo {author} {\bibfnamefont {A.}~\bibnamefont
  {Korm\'anyos}}, \bibinfo {author} {\bibfnamefont {V.}~\bibnamefont
  {Z\'olyomi}}, \bibinfo {author} {\bibfnamefont {N.~D.}\ \bibnamefont
  {Drummond}}, \bibinfo {author} {\bibfnamefont {P.}~\bibnamefont {Rakyta}},
  \bibinfo {author} {\bibfnamefont {G.}~\bibnamefont {Burkard}}, \ and\
  \bibinfo {author} {\bibfnamefont {V.~I.}\ \bibnamefont {Fal'ko}},\ }\bibfield  {title} {\bibinfo {title} {Monolayer MoS${}_{2}$: Trigonal warping, the $\ensuremath{\Gamma}$ valley, and spin-orbit coupling effects}}, \href
  {\doibase 10.1103/PhysRevB.88.045416} {\bibfield  {journal} {\bibinfo
  {journal} {Phys. Rev. B}\ }\textbf {\bibinfo {volume} {88}},\ \bibinfo
  {pages} {045416} (\bibinfo {year} {2013})}\BibitemShut {NoStop}%
\bibitem [{\citenamefont {Battilomo}\ \emph {et~al.}(2019)\citenamefont
  {Battilomo}, \citenamefont {Scopigno},\ and\ \citenamefont
  {Ortix}}]{battilomo2019}%
  \BibitemOpen
  \bibfield  {author} {\bibinfo {author} {\bibfnamefont {R.}~\bibnamefont
  {Battilomo}}, \bibinfo {author} {\bibfnamefont {N.}~\bibnamefont {Scopigno}},
  \ and\ \bibinfo {author} {\bibfnamefont {C.}~\bibnamefont {Ortix}},\ }\bibfield  {title} {\bibinfo {title} {Berry Curvature Dipole in Strained Graphene: A Fermi Surface Warping Effect}}, \href
  {\doibase 10.1103/PhysRevLett.123.196403} {\bibfield  {journal} {\bibinfo
  {journal} {Phys. Rev. Lett.}\ }\textbf {\bibinfo {volume} {123}},\ \bibinfo
  {pages} {196403} (\bibinfo {year} {2019})}\BibitemShut {NoStop}%
\bibitem [{\citenamefont {Wakatsuki}\ \emph {et~al.}(2017)\citenamefont
  {Wakatsuki}, \citenamefont {Saito}, \citenamefont {Hoshino}, \citenamefont
  {Itahashi}, \citenamefont {Ideue}, \citenamefont {Ezawa}, \citenamefont
  {Iwasa},\ and\ \citenamefont {Nagaosa}}]{wakatsuki2017}%
  \BibitemOpen
  \bibfield  {author} {\bibinfo {author} {\bibfnamefont {R.}~\bibnamefont
  {Wakatsuki}}, \bibinfo {author} {\bibfnamefont {Y.}~\bibnamefont {Saito}},
  \bibinfo {author} {\bibfnamefont {S.}~\bibnamefont {Hoshino}}, \bibinfo
  {author} {\bibfnamefont {Y.~M.}\ \bibnamefont {Itahashi}}, \bibinfo {author}
  {\bibfnamefont {T.}~\bibnamefont {Ideue}}, \bibinfo {author} {\bibfnamefont
  {M.}~\bibnamefont {Ezawa}}, \bibinfo {author} {\bibfnamefont
  {Y.}~\bibnamefont {Iwasa}}, \ and\ \bibinfo {author} {\bibfnamefont
  {N.}~\bibnamefont {Nagaosa}},\ }\bibfield  {title} {\bibinfo {title} {Nonreciprocal charge transport in noncentrosymmetric superconductors}}, \href {\doibase 10.1126/sciadv.1602390}
  {\bibfield  {journal} {\bibinfo  {journal} {Sci. Adv.}\ }\textbf {\bibinfo
  {volume} {3}},\ \bibinfo {pages} {e1602390} (\bibinfo {year}
  {2017})}\BibitemShut {NoStop}%
\bibitem [{\citenamefont {Wakatsuki}\ and\ \citenamefont
  {Nagaosa}(2018)}]{wakatsuki2018}%
  \BibitemOpen
  \bibfield  {author} {\bibinfo {author} {\bibfnamefont {R.}~\bibnamefont
  {Wakatsuki}}\ and\ \bibinfo {author} {\bibfnamefont {N.}~\bibnamefont
  {Nagaosa}},\ }\bibfield  {title} {\bibinfo {title} {Nonreciprocal Current in Noncentrosymmetric Rashba Superconductors}}, \href {\doibase 10.1103/PhysRevLett.121.026601} {\bibfield
  {journal} {\bibinfo  {journal} {Phys. Rev. Lett.}\ }\textbf {\bibinfo
  {volume} {121}},\ \bibinfo {pages} {026601} (\bibinfo {year}
  {2018})}\BibitemShut {NoStop}%
\bibitem [{\citenamefont {Hoshino}\ \emph {et~al.}(2018)\citenamefont
  {Hoshino}, \citenamefont {Wakatsuki}, \citenamefont {Hamamoto},\ and\
  \citenamefont {Nagaosa}}]{hoshino2018}%
  \BibitemOpen
  \bibfield  {author} {\bibinfo {author} {\bibfnamefont {S.}~\bibnamefont
  {Hoshino}}, \bibinfo {author} {\bibfnamefont {R.}~\bibnamefont {Wakatsuki}},
  \bibinfo {author} {\bibfnamefont {K.}~\bibnamefont {Hamamoto}}, \ and\
  \bibinfo {author} {\bibfnamefont {N.}~\bibnamefont {Nagaosa}},\ }\bibfield  {title} {\bibinfo {title} {Nonreciprocal charge transport in two-dimensional noncentrosymmetric superconductors}}, \href
  {\doibase 10.1103/PhysRevB.98.054510} {\bibfield  {journal} {\bibinfo
  {journal} {Phys. Rev. B}\ }\textbf {\bibinfo {volume} {98}},\ \bibinfo
  {pages} {054510} (\bibinfo {year} {2018})}\BibitemShut {NoStop}%
\bibitem [{\citenamefont {Ideue}\ \emph {et~al.}(2017)\citenamefont {Ideue},
  \citenamefont {Hamamoto}, \citenamefont {Koshikawa}, \citenamefont {Ezawa},
  \citenamefont {Shimizu}, \citenamefont {Kaneko}, \citenamefont {Tokura},
  \citenamefont {Nagaosa},\ and\ \citenamefont {Iwasa}}]{ideue2017}%
  \BibitemOpen
  \bibfield  {author} {\bibinfo {author} {\bibfnamefont {T.}~\bibnamefont
  {Ideue}}, \bibinfo {author} {\bibfnamefont {K.}~\bibnamefont {Hamamoto}},
  \bibinfo {author} {\bibfnamefont {S.}~\bibnamefont {Koshikawa}}, \bibinfo
  {author} {\bibfnamefont {M.}~\bibnamefont {Ezawa}}, \bibinfo {author}
  {\bibfnamefont {S.}~\bibnamefont {Shimizu}}, \bibinfo {author} {\bibfnamefont
  {Y.}~\bibnamefont {Kaneko}}, \bibinfo {author} {\bibfnamefont
  {Y.}~\bibnamefont {Tokura}}, \bibinfo {author} {\bibfnamefont
  {N.}~\bibnamefont {Nagaosa}}, \ and\ \bibinfo {author} {\bibfnamefont
  {Y.}~\bibnamefont {Iwasa}},\ }\bibfield  {title} {\bibinfo {title} {Bulk rectification effect in a polar semiconductor}}, \href {\doibase 10.1038/nphys4056} {\bibfield
  {journal} {\bibinfo  {journal} {Nat. Phys.}\ }\textbf {\bibinfo {volume}
  {13}},\ \bibinfo {pages} {578} (\bibinfo {year} {2017})}\BibitemShut
  {NoStop}%
\bibitem [{\citenamefont {Asgari}\ and\ \citenamefont
  {Culcer}(2022)}]{asgari2022}%
  \BibitemOpen
  \bibfield  {author} {\bibinfo {author} {\bibfnamefont {R.}~\bibnamefont
  {Asgari}}\ and\ \bibinfo {author} {\bibfnamefont {D.}~\bibnamefont
  {Culcer}},\ }\bibfield  {title} {\bibinfo {title} {Unidirectional valley-contrasting photocurrent in strained transition metal dichalcogenide monolayers}}, \href {\doibase 10.1103/PhysRevB.105.195418} {\bibfield
  {journal} {\bibinfo  {journal} {Phys. Rev. B}\ }\textbf {\bibinfo {volume}
  {105}},\ \bibinfo {pages} {195418} (\bibinfo {year} {2022})}\BibitemShut
  {NoStop}%
\bibitem [{tru()}]{truell1969}%
  \BibitemOpen
  \href@noop {} {}\bibinfo {note} {R. Truell, C. Elbaum, and B. B. Chick,
  \emph{Ultrasonic Methods in Solid State Physics} (Academic Press, New York,
  1969).}\BibitemShut {Stop}%
\bibitem [{lut()}]{luthi2004}%
  \BibitemOpen
  \href@noop {} {}\bibinfo {note} {B. L\"uthi, \emph{Physical Acoustics in the
  Solid State} (Springer-Verlag, Berlin, 2004).}\BibitemShut {Stop}%
\end{thebibliography}

%

\end{document}